\documentclass[12pt,preprint]{aastex} 
\usepackage{psfig,natbib,amsmath}

\newcommand{\citepeg}[1]{\citep[{e.g.,}][]{#1}}

\def\lsim{\hbox{ \rlap{\raise 0.425ex\hbox{$<$}}\lower 0.65ex\hbox{$\sim$}}}
\def\gsim{\hbox{ \rlap{\raise 0.425ex\hbox{$>$}}\lower 0.65ex\hbox{$\sim$}}}
\def\arcmin{\hbox{$^\prime$}}
\def\arcsec{\hbox{$^{\prime\prime}$}}
\def\arcdeg{\mbox{$^\circ$}}%

\def\fh{\hbox{$~\!\!^{\rm h}$}}
\def\fm{\hbox{$~\!\!^{\rm m}$}}
\def\fs{\hbox{$~\!\!^{\rm s}$}}
\def\ale{\mathrel{\hbox{\rlap{\hbox{\lower4pt\hbox{$\sim$}}}\hbox{$<$}}}}
\def\age{\mathrel{\hbox{\rlap{\hbox{\lower4pt\hbox{$\sim$}}}\hbox{$>$}}}}

\def\kms{km~s$^{-1}$ }
\def\swift{{\textit{Swift}}\,}
\def\thegrb{GRB\,080319B }

\begin{document}

\title{Observations of the Naked-Eye GRB 080319B: Implications of Nature's
Brightest Explosion}

\def\berk{1}
\def\sloan{2}
\def\chicago{3}
\def\lcogt{4}
\def\tautenberg{5}
\def\lick{6}
\def\uva{7}
\def\das{8}
\def\york{9}
\def\swin{10}
\def\cfa{11}

\author{J.~S. Bloom\altaffilmark{\berk,\sloan},
        D.~A. Perley\altaffilmark{\berk},
        W. Li\altaffilmark{\berk}, \\
        N.~R. Butler\altaffilmark{\berk},
        A.~A. Miller\altaffilmark{\berk},
        D. Kocevski\altaffilmark{\berk},
        D.~A. Kann\altaffilmark{\tautenberg},\\
        R.~J. Foley\altaffilmark{\berk},
        H.-W. Chen\altaffilmark{\chicago},
        A.~V. Filippenko\altaffilmark{\berk}, 
        D.~L. Starr\altaffilmark{\berk, \lcogt}, \\
        B. Macomber\altaffilmark{\berk},
        J.~X. Prochaska\altaffilmark{\lick},
        R. Chornock\altaffilmark{\berk}, \\
		D. Poznanski\altaffilmark{\berk}, 
		S. Klose\altaffilmark{\tautenberg},
		M. F. Skrutskie\altaffilmark{\uva}, \\
		S. Lopez\altaffilmark{\das}, P. Hall\altaffilmark{\york}, 
		K. Glazebrook\altaffilmark{\swin}, 
		and C. H. Blake\altaffilmark{\cfa}}

\affil{$^\berk$ Department of Astronomy, 
        University of California, Berkeley, CA 94720-3411.}

\affil{$^\sloan$ Sloan Research Fellow.}

\affil{$^\chicago$ Department of Astronomy \& Astrophysics, University of
Chicago, Chicago, IL 60637.}

\affil{$^\lcogt$ Las Cumbres Global Telescope Network, 6740 Cortona Dr. Santa
Barbara, CA 93117.}

\affil{$^\tautenberg$ Th\"uringer Landessternwarte Tautenburg, Sternwarte 5,
D-07778 Tautenburg, Germany.}

\affil{$^\lick$ University of California Observatories/Lick
Observatory, University of California, Santa Cruz, CA 95064.}

\affil{$^\uva$ Department of Astronomy, P.O. Box 3818, University of
Virginia, Charlottesville, VA 22903-0818.}

\affil{$^\das$ 
Departamento de Astronomía, Universidad de Chile, Casilla 36-D, Santiago, Chile.}

\affil{$^\york$ Physics and Astronomy, Toronto, Ontario, M3J 1P3, Canada.}

\affil{$^{\swin}$ Centre for Astrophysics and Supercomputing, Swinburne University of Technology, Hawthorn, VIC 3122, Australia}

\affil{$^{\cfa}$ Harvard-Smithsonian Center for Astrophysics, 60 Garden Street, Cambridge,
MA 02138.}

\begin{abstract}
The first gamma-ray burst (GRB) confirmed to be bright enough to be
seen with the naked eye, GRB~080319B at redshift $z = 0.937$, allowed
for exquisite follow-up observations across the electromagnetic
spectrum. We present our detailed optical and infrared observations of
the afterglow, consisting of over 5000 images starting 51~s after the
GRB trigger, in concert with our own analysis of the {\it Swift} UVOT,
BAT, and XRT data. The event is extreme not only in observed
properties but intrinsically: it was the most luminous event ever recorded
at optical and infrared wavelengths and had an exceedingly high
isotropic-equivalent energy release in $\gamma$-rays. At early times,
the afterglow evolution is broadly consistent with being reverse-shock
dominated, but then is subsumed by a forward shock at around 1000~s.
The overall spectral energy distribution, spanning from ultraviolet
through near-infrared wavelengths, shows no evidence for a significant
amount of dust extinction in the host frame.  The afterglow evolution,
however, is highly chromatic: starting at about 1000~s the index
shifts blueward before shifting back to the red at late times.  In our
deepest late-time observations, we find tentative evidence for an
optical jet break and a luminous supernova.  Finally, we examine
the detectability of such events with current and future facilities
and find that such an event could be detected in gamma-rays by BAT out
to $z$ = 10.7 ($8\sigma$), while the nominal EXIST sensitivity would
allow detection to $z \approx 32$. At $K$ band, this source would have been
easily detected with meter-class telescopes to $z \approx 17$.
\end{abstract}


\section{Introduction}

The longevity and burst-discovery prowess of the \swift mission
\citep{Gehrels:2004p672} has led to a boom in correlative studies of
the properties of a large sample of gamma-ray bursts (GRBs) observed
systematically and uniformly. At the same time, the sheer number of
bursts encompasses an ever-expanding volume of interesting parameter
space, often revealing rarities that help to redefine and shape the
totality of our understanding of the phenomenon. It is in the backdrop of the latter
 that \thegrb superlatively reigns.

At 06:12:49 (UTC is used throughout this paper), \thegrb
triggered \citep{rgh+08} the {\swift} Burst Alert Telescope (BAT), the
second GRB trigger that day in what would be 5 GRB triggers in a
24\,hr period.  The extreme brightness of the burst at high energies,
in the X-ray afterglow, and in the ultraviolet/optical/infrared
(UV/O/IR) afterglow led to a flurry of follow-up observations, many
automatically triggered on robotic facilities. Contemporaneous
imaging, both all-sky and directed, uncovered a fast-rising optical
afterglow which peaked a time from trigger of $t \approx 18.3$~s at
$V \approx 5.3$~mag
\citep{Cwiok08_GCN7445,Covino08_GCN7446,Swan08_GCN7470,Karpov08_GCN7452,Schubel08_GCN7461,Wozniak08_GCN7464};
this makes the afterglow of \thegrb the first confirmed counterpart
that could have been seen with the unaided eye in dark skies.

At a redshift of $z=0.937$ \citep{Vreeswijk08}, the event was
relatively nearby compared to the \swift distribution of long-duration
events
\citepeg{2006A&A...447..897J,2006MNRAS.372.1034D,2007ApJ...661..394L},
yet near the median for redshifts of pre-\swift events. One aim of
this article is to place the observed properties (particularly
energetics) in the context of both distributions; we show in \S
\ref{sec:als} that \thegrb was not only potentially the
highest-fluence event ever observed, it also had an
isotropic-equivalent energy release comparable to the highest known
values yet recorded.  The other significant aim is to analyze in
detail our long-wavelength observations, in both the optical and IR,
with high time cadence on moderate-sized robotic telescopes. The
quality of the data, coupled with those available in the literature
and from public archives, reveals a complex evolution of the afterglow
that we attempt to reconcile with canonical afterglow theory.  Unless
noted, we assume a concordance cosmology with $H_0 = 71$ \kms
Mpc$^{-1}$, $\Omega_\Lambda = 0.70$, and $\Omega_m = 0.3$. A redshift
of $z=0.937$ corresponds to a luminosity distance of $6011.3$~Mpc
(distance modulus 43.89~mag). All of the results presented herein,
though generally consistent with our previous results in GCN
Circulars\footnote{\url{http://gcn.gsfc.nasa.gov/ .}}, supersede them.

\section{Optical/IR Observations and Reductions}
\label{sec:obs}

\begin{figure*}[p] 
\centerline{\psfig{file=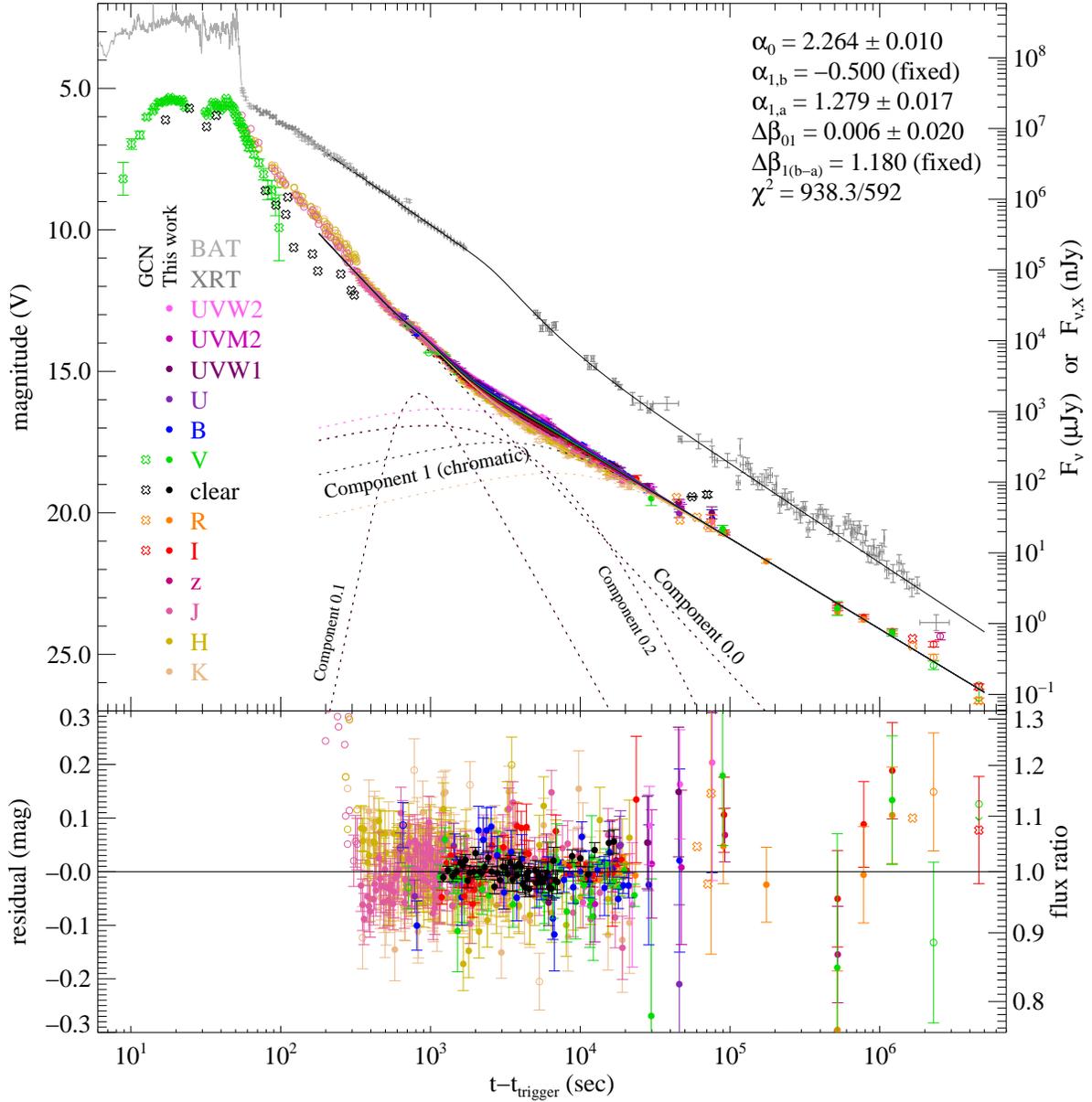,width=6.5in,angle=0}}
\caption[] {\small Light curves of the \thegrb long-wavelength
afterglow, fit by our empirical model, which allows (and in this case
prefers) color change. This is a combination of data from the GCN
Circulars ($\times$ symbols, including the prompt light curve as
plotted by \citealt{Karpov08_GCN7558}, in green), our observations
from various ground-based instruments (KAIT, the Lick Nickel 1m, and
PAIRITEL) and our re-reductions of the \swift UVOT, XRT, and BAT data.
The afterglow decays extremely rapidly, dropping from mag 5 to 21 in
less than one day. For clarity, UV/O/IR data are corrected to $V$
[Vega] mag using the model. Individual broken power-law components are shown as dotted lines; these are summed together to generate the fitted model (solid line). Different colors indicate different filters. Empty points were not used in the fitting
in \S \ref{sec:canonball}.}
\label{fig:lc}
\end{figure*}

\subsection{PAIRITEL}

The Peters Automated Infrared Imaging Telescope (PAIRITEL;
\citealt{Bloom:2006p4665}), responding automatically to the {\swift}
trigger via an open-source
package\footnote{\url{http://sourceforge.net/projects/pygcnsock .}}
connected to the GCN, began taking data on the field at 06:13:39.7 ($t =
51$~s after the \swift trigger). The 1.3~m telescope is equipped with NICMOS3 arrays to
simultaneously observe in bands {\em J}, {\em H}, and {\em K$_s$}
(1.2, 1.6, and 2.2~$\mu$m, respectively).  Each image consists of a
256 $\times$ 256 pixel array with a scale of $2''$
pixel$^{-1}$. Data are obtained as a double-correlated read, with a 51~ms 
accumulation since reset on the first (``short'') read and 7.851~s on the
second (``long'') read. Both the long and the short read exposures are
saved to disk. The telescope is dithered by $\sim 1'$ after every
third exposure. With no shutter, the standard reductions necessarily
incorporate per-pixel models of the combined detector dark current and
sky flux fitted over time \citep{wfb+08}. In a modified version of our
automated pipeline, once these time-specific frames (``sky+dark'')
are subtracted from the object frames, they are combined into 3$-$12
file intermediate mosaics with effective integration times of 24$-$96~s per
image, which in turn are stacked to form the final mosaics.

During the first few minutes on target (253~s in $J$, 289~s in $H$,
and 289~s in $K_s$), the afterglow was saturated in
the exposures, where we have adopted the saturation/non-linearity threshold
determined from data for the 2MASS South camera\footnote{{\tt
http://www.ipac.caltech.edu/2mass/releases/allsky/ \\ doc/sec4\_2.html
.}}: 37000 counts in a single pixel in the $J$ and $H$ bands, and 33000
counts in a single pixel in the $K_s$ band. In addition, a number of the
final science frames suffered from poor sky+dark-frame subtraction and
could not be photometered or included in the final mosaics. These
frames were removed following an inspection by eye prior to the
construction of the final mosaics. Less than 4\% of the 7.8~s
exposures were removed following this procedure.

\begin{figure*}[p] 
\centerline{\psfig{file=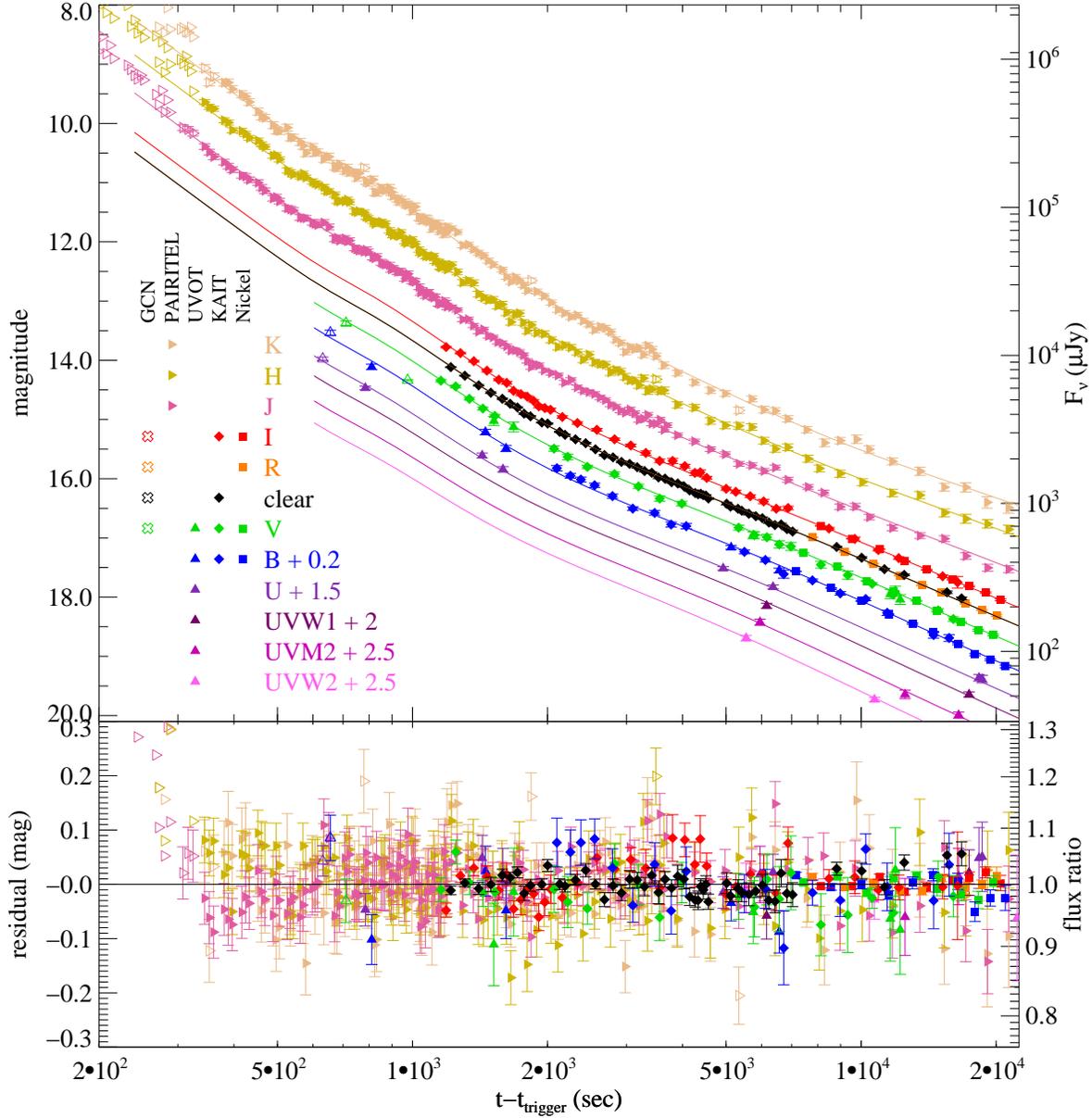,width=6.5in,angle=0}}
\caption[] {\small Detailed optical-IR light curve covering much of the first
night after the afterglow, demonstrating the exquisite time sampling and
overall goodness-of-fit of the model.  The earliest-time PAIRITEL observations
are saturated/non-linear and uncertain (see \S \ref{sec:nl}); these are not included in fitting or modeling.}
\label{fig:lczoom}
\end{figure*}

Our initial IR photometry report \citep{bsp08} noted that the
afterglow had been strongly detected simultaneously in all three
filters during the first few minutes of observations.  Indeed, the
ability to detect the transient in single 7.8~s exposures lasted for a
few hours.  In total, we obtained 1822 simultaneous $J$, $H$, and
$K_{s}$ observations of the GRB (5466 total frames) over the course of
6 hr, though as the afterglow begins to fade we bin together
individual frames to increase the signal-to-noise ratio and improve
readability of the light curve.

\subsubsection{Photometry on Unsaturated Frames}
\label{sec:nl}

A custom pipeline was used to perform photometry on the 7.8 s exposures constructed from the difference  of the  two readouts which occurred 7.851 and 0.051 s
   following array reset. This pipeline was also used on the  intermediate
mosaics constructed from the 7.8s frames. Zeropoint determinations were made in comparison to the 2MASS
catalog. The statistical uncertainties are small, while the total
uncertainty on any individual image is dominated by systematics,
especially at early times, associated with our ability to accurately
zeropoint relative to 2MASS. We quantify this systematic uncertainty
empirically by measuring the photometric scatter in a mag 10 star
(hereafter ``S1''), located at $\alpha$ = 14\fh32\fm03\fs.55, $\delta$
= +36\arcdeg18\arcmin29\arcsec.6 (J2000), a few arcmin from the
position of the GRB optical afterglow (OA). In all three bands the systematic contribution
is $\sim$4\%. The resulting light curves are presented in
Figs. \ref{fig:lc}--\ref{fig:lczoom} and listed in Table
\ref{tab:photlogptel}.

\subsubsection{Photometry on Early-Time Imaging}

Due to the extreme brightness of the early afterglow, a large number
of early PAIRITEL frames are highly saturated or otherwise nonlinear (when the magnitude is $\lesssim$ 9).
We employed two methods to recover the photometry in these early epochs (see Tables \ref{tab:shortr}--\ref{tab:annulus}). First, since there is sufficient signal in the  51 ms short reads to detect the GRB well above the noise when $JHK_s$ $\gtrsim$ 7, we use the 
procedure described in \citet{2007ApJ...669.1072E} to extract the flux from the GRB 
in the short reads and calibrate this to the 2MASS system using stars in 
the corresponding long reads. The uncertainty of these measurements is 
primarily dominated by the determination of the zeropoint. Second, we developed a simple method to extract photometry from saturated ``long minus short'' frames, using an annulus centered on the GRB position. We
use S1 to determine the zeropoint for each of the individual saturated
long minus short (7.8~s) frames. The inner annulus radius was chosen
to exclude any saturated pixels in any frame, while the outer annulus
radius was selected to provide sufficient signal without including too
much noise as the wings of the point-spread function (PSF) become
dominated by noise from the background.  The accuracy of these early
measurements of the GRB afterglow is strongly limited by the determination of
the zeropoint: high near-IR backgrounds and the annulus aperture lead
to a large scatter in the determined zeropoints for each frame. Thus,
the uncertainty for these early measurements is dominated by a large
systematic term, defined as the scatter of the individual zeropoints.
We include these points in early-time light-curve plots and in our
data table to show the general early behavior of the IR afterglow, but
do not include them in any fitting or model analysis.  We emphasize that photometry of
these saturated frames is subject to unaccounted-for systematic uncertainties, but note qualitative agreement with other submitted early IR imaging \citep{Racusineteveryone}.

\subsection{KAIT and Nickel}

At Lick Observatory, the Katzman Automatic Imaging Telescope (KAIT;
\citealt{Filippenko2001}) GRB alert system \citep{Li:2003p703}
responded to \thegrb automatically. While ordinarily the time to slew
is $<1$~min, since KAIT was following GRB\,080319A (which triggered
{\swift} 30 min earlier) our first observations of \thegrb did not
start until 19 min after the BAT trigger. We followed the GRB OA with
a combination of filters ($BVI$ and unfiltered) and with varying
exposure times (20~s initially, then 40~s, then 300~s guided
exposures), and continued until 280 min after the BAT trigger. We also
observed \thegrb with the Lick 1-m Nickel telescope remotely from the
University of California, Berkeley, between 116 and 392 min after the
BAT trigger, using $BVRI$ filters and exposure times of 300~s and
360~s.

To reduce the KAIT and Nickel data, we used the PSF-fitting technique
in IRAF/DAOPHOT. Instrumental magnitudes were measured for the GRB OA
and several local standard stars, and conversion to the standard
$BVRI$ system was accomplished using the SDSS calibration
\citep{ceh+08,DR6} of the field (the magnitudes in the SDSS magnitude
system were converted to $BVRI$ following the recipe of Lupton
2005\footnote{http://www.sdss.org/dr6/algorithms/ \\
sdssUBVRITransform.html\#Lupton2005
.}).  Two of the SDSS standards in our field are shared with the
standard star calibration of \cite{Henden08_GCN7528}; because the
posted magnitudes for both stars are consistent within the
uncertainties, we adopted the SDSS system due to the much larger
number of stars and the ability to extend to deep late-time
observations.  The final KAIT and Nickel photometry is listed in
Tables \ref{tab:photlogkait} and \ref{tab:photlognickel}.  A list
of the converted $BVRI$ magnitudes of the bright standards used
in this calibration is given in Table \ref{tab:lickstars}.

\subsection{UVOT}

To extend the wavelength coverage, we downloaded the \swift UVOT data
from the quicklook data archive. The Level 2 sky image data in $U$,
$B$, and $V$ were analyzed according to the photometry calibration and
recipe by \citet{Li:2006p4490}. We also followed the procedure
reported in \citet{Poole2008}, and found that when the GRB was bright,
the two procedures yield similar results. When the GRB became faint,
the Li et al. (2006) procedure yields measurements with smaller
uncertainties and better overall agreement with the KAIT and Nickel
data in the $B$ and $V$ passbands, likely due to the smaller adopted
photometry aperture. The \swift UV filters ($UVW1$, $UVM1$, and
$UVW2$) were reduced following \citet{Poole2008}.  The final UVOT
photometry is reported in Table \ref{tab:photloguvot}.

\subsection{Gemini Spectroscopy}

An optical spectrum of \thegrb was obtained under program
GS-2008A-Q-20 beginning at 08:23 on 19 Mar. 2008 \citep{Foley08} using
the Gemini~South 8-m telescope with GMOS \citep{Hook04}.  We used a
slit of width $0.75''$, the R831 grating, and a OG515 filter.  Two
1800~s exposures were obtained with slightly different central
wavelengths of 7000 and 7100~\AA\ and read out in $2\times 2$
binning. Standard CCD processing and spectrum extraction were
accomplished with IRAF \citep[for more details, see][]{Foley06}.  The
data were extracted using the optimal algorithm of
\citet{Horne86}. The spectrum shows a featureless continuum with no
strong absorption systems, emission features, or spectral breaks.   Unfortunately, our wavelength range does not cover the spectral range of the VLT spectrum where strong absorption features were seen \citep{Vreeswijk08}. 
With our spectral coverage, and given the proposed redshift of GRB~080319B 
\citep{Vreeswijk08}, the strongest ISM absorption lines for
a galaxy at $z=0.937$ are expected to be \ion{Ti}{2}~3384 and the 
\ion{Ca}{2} doublet at $\lambda \approx 3950$\AA.  
We do not detect \ion{Ti}{2} $\lambda 3384$ to a
3$\sigma$ rest-frame absorption equivalent width limit of $W=0.05$
\AA\ over a spectral resolution element, while the \ion{Ca}{2} H \&
K absorption doublet is blended with the atmospheric A-band
absorption. 
No additional features were found in our observed 
wavelength range of 5950--8150~\AA.

\subsection{Gemini Imaging}

In the nights following the burst, we began a program of additional
imaging using Gemini~South and Gemini~North.  On the first night
following the event, we acquired 4 $\times$ 180~s of GMOS imaging in
each of the $g$, $r$, $i$, and $z$ filters under excellent seeing
conditions (0.85\arcsec) despite the low elevation of the target
(airmass of 2.5).  A second epoch was acquired the following night in
$r$ only, and additional multicolor epochs were taken later on March
25 and 28, and April 2.  Imaging was taken at Gemini~North on April 14
($g$, $r$, $i$) and 17 ($z$).
Data were reduced using the standard
Gemini IRAF package and photometered using SExtractor
\citep{SExtractor} aperture magnitudes.  The field was calibrated
relative to select stars from SDSS DR6 (\citealt{DR6}, Table
\ref{tab:sdssstars}).  The Gemini filters appear to be somewhat
nonstandard compared to the SDSS survey filters, so significant
color-term corrections were necessary.  As the number of comparison
stars used was large we were able to calculate these individually for
each observation.  The color dependence is about 20\% in $g$ on
Gemini~North and 30\% on Gemini~South.  In $r$, $i$, and $z$ it is
about 10\%, 15\%, and 5\% (respectively) on both telescopes.  The
final photometry is given in Table \ref{tab:photloggemini}.  For use
in the light-curve fitting (for which a comparison to the early-time
$BVRI$ observations is necessary), the $gri$ magnitudes for the
afterglow were then converted back to $VRI$ using the equations of
Lupton (2005).

\subsection{GCN Circulars}

Finally, to supplement our data at very early times and late times, we
downloaded additional photometry from the GCN Circulars.  Of
particular note, we downloaded the corrected TORTORA light curve
\citep{Karpov08_GCN7558} containing high-quality optical photometry
throughout the prompt phase, and {\it Hubble Space Telescope (HST)}
data points at late times \citep{Tanvir08_GCN7569,Levan08_GCN7710},
which allow us to complete the entire light curve of this event.  Due
to the fact that GCN observations are preliminary and may have large
calibration offsets, we do not use any of the GCN points in any of our
fits or models, though we do show them in light-curve plots.  Instead,
we focus our analysis on the intermediate phase of the burst,
from 300~s to $10^6$~s, during which we have good time sampling.
The GCN data shown in our plots are listed in Table
\ref{tab:photloggcn}.

\subsection{Swift BAT and XRT}
\label{sec:hep}

Our high-energy reduction pipeline is described in detail by
\citet{Butler:2007p4518} for the {\swift} BAT and by
\citet{Butler:2007p4519} for the {\swift} XRT. \thegrb exhibits one
dominant emission episode of duration $\Delta t \approx 60$~s,
composed of multiple unresolved spikes.  The BAT spectrum in the time
interval $-$1.1~s to 57.4~s is acceptably fit ($\chi^2/\nu=13.55/55$)
by a simple power law with photon index $\alpha=-1.01\pm0.02$ and
energy fluence $(1.96\pm0.03) \times 10^{-4}$~erg cm$^{-2}$
(15--350~keV). The BAT catches only the low-energy portion of a
spectrum extending beyond $E_{\rm peak,obs} = 651^{+13}_{-14}$~keV
\citep{gam+08}, with an isotropic-equivalent energy release in
$\gamma$-rays of $\sim 10^{54}$~erg.  There is evidence for a
gamma-ray tail detected up to $\sim1000$~s as seen in Fig.\
\ref{fig:lc}. This extended emission at late times is similar to that
observed in the extremely bright BATSE GRB 980923 \citep{Giblin99} and
\swift GRB 061007 \citep{Mundell07,Schady07}.
  
The {\swift} XRT began observing during the tail of the prompt
emission phase at $t=66$~s.  We find (see, also, \citealt{but08}) that
the X-ray spectrum is unchanging until quite late times ($t = 2.91$~Ms)
despite a break in the X-ray light curve at $t \approx 1$~ks (Fig.\
\ref{fig:lc}).  The combined XRT/BAT data are well fit by an absorbed
power law with $\Gamma=1.814\pm0.011$ and an excess column density
over Galactic of $N_{\rm H} = (1.87\pm 0.13) \times 10^{21}$~cm$^{-2}$
at $z=0.937$.  The early-time windowed timing (WT) mode (66 s $<t<$ 4.95 ks) X-ray
photon index $\Gamma_1=1.814\pm 0.013$ is closely consistent with the
late-time photon-counting (PC) mode (4.95 ks $<t<$ 2.91 Ms) X-ray photon index
$\Gamma_2=1.80\pm 0.04$, assuming a constant $N_{\rm H}$.  From
negligible $\lesssim 1$\% variations in the X-ray hardness ratio
\citep[see, e.g.,][]{bnk07b}, the magnitude of any secular trends in
$\Gamma_1$ or $\Gamma_2$ must be at the few-percent level or less, and
there is no evidence for variation in $N_{\rm H}$.

\begin{figure*}[p] 
\centerline{\psfig{file=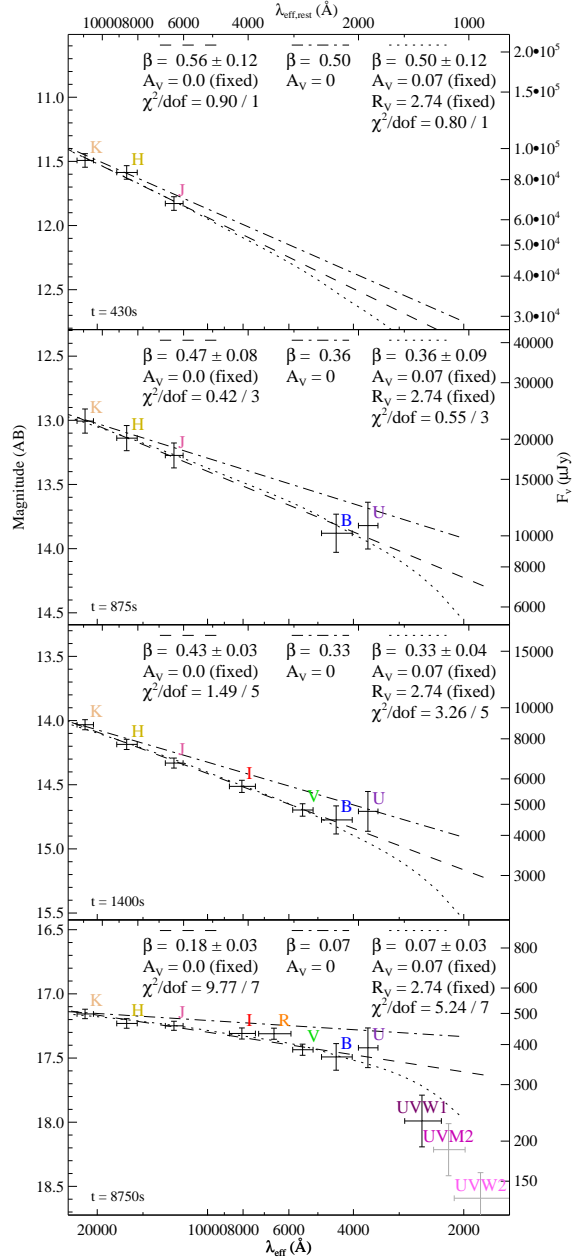,width=3.0in,angle=0}}
\caption[] {\small Photometric spectral energy distributions (SEDs)
generated over specific time ranges during the first night after the
burst from combined PAIRITEL, KAIT, Nickel, and UVOT measurements.
All plots have the same horizontal and vertical scale, differing only
in the vertical offset.  The assumed host-galaxy extinction is based
on a fit using a combined SED from all epochs (a fit assuming no host
extinction is also shown for comparison).  A clear red-to-blue
transition is evident during this time. The three lines show the various fits to the data with the relevant parameters noted. ``Fixed'' indicates that that parameter is held constant in the fit.}
\label{fig:seds}
\end{figure*}

\begin{figure*}[p] 
\centerline{\psfig{file=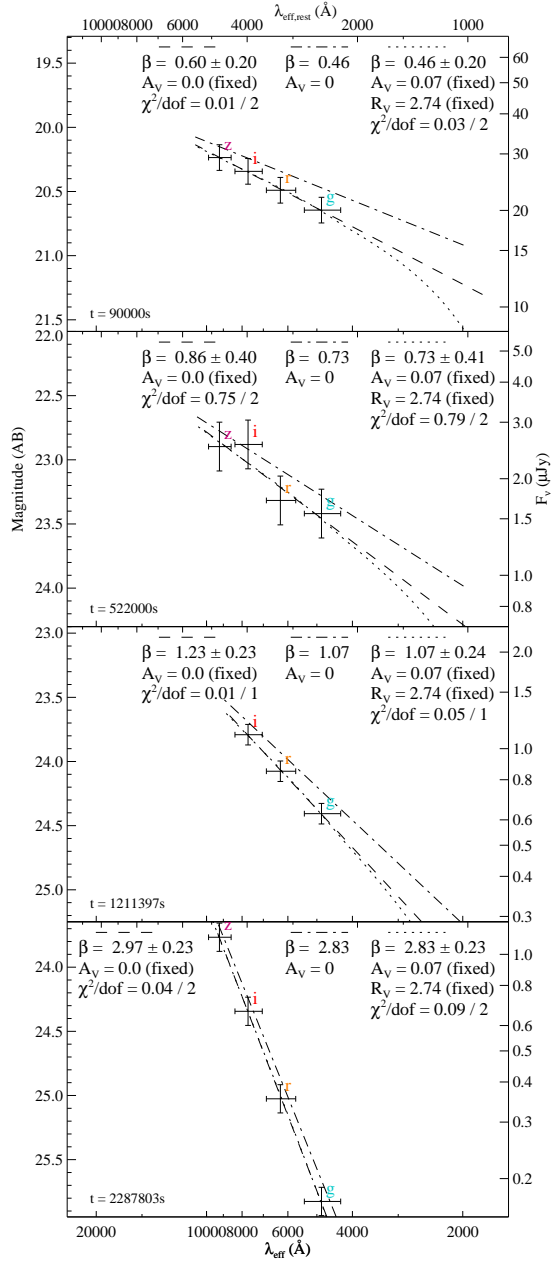,width=3.0in,angle=0}}
\caption[] {\small Photometric SEDs generated on subsequent nights
from multicolor Gemini~South and Gemini~North observations.  The scale
and assumed host extinction are the same as in Figure \ref{fig:seds}.
The color is consistent with constant evolution early, but shifts
dramatically redward at very late times, likely due to the appearance
of a very luminous supernova.}
\label{fig:seds2}
\end{figure*}

\begin{figure*}[p] 
\centerline{\psfig{file=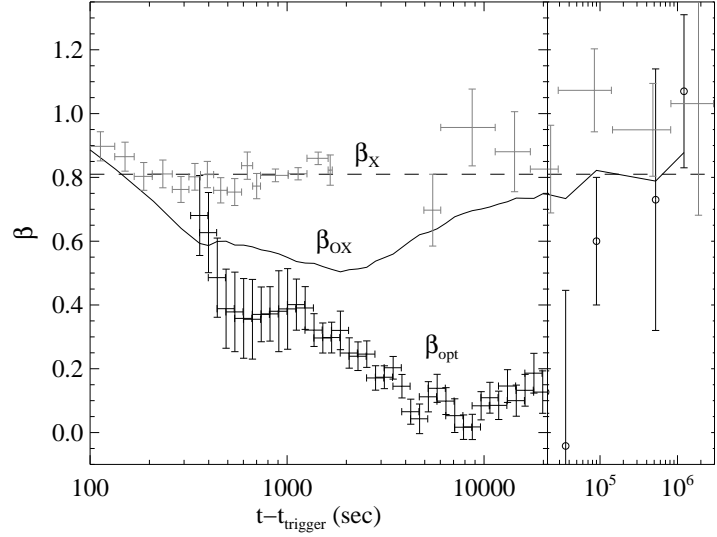,width=3.8in,angle=0}}
\caption[] {\small Evolution of the afterglow spectral index $\beta$ with time. The optical spectral index $\beta_{\rm opt}$ is measured from a fit to the UV-optical-IR data from each of 40 different overlapping bins ranging from 360 s to 22000 s, plus one additional series of UVOT exposures and the first three multicolor Gemini epochs. The X-ray spectral index is also plotted (gray points), with the dashed line showing the best-fit value assuming no spectral evolution (which the data are consistent with).  The optical-to-X-ray spectral index $\beta_{\rm OX}$ is defined as the index between the X-ray (normalized at 1 keV) and the $V$-band optical flux. At early times the optical spectral index evolves from red to blue to red again. The X-ray, optical, and optical-to-X-ray spectral indices are all consistent at late times, which may suggest that both optical and X-ray bands are in the same synchrotron regime.}
\label{fig:beta}
\end{figure*}

\begin{figure*}[p] 
\centerline{\psfig{file=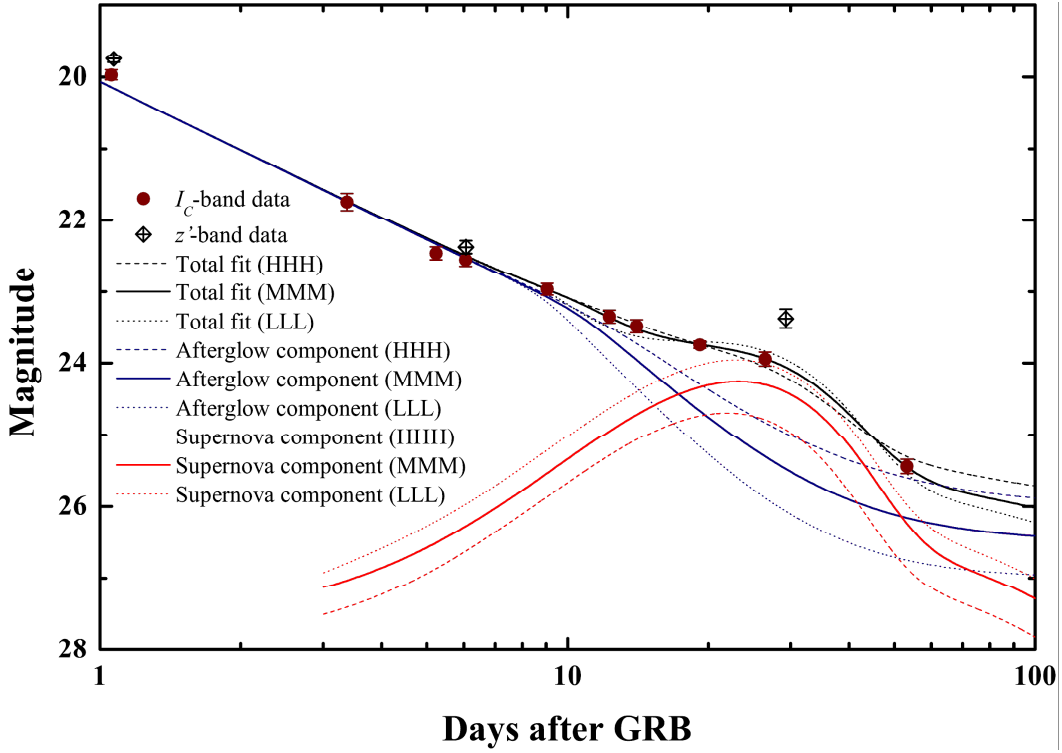,width=5.5in,angle=0}}
\caption[] {\small Fits to the late-time $I_C$-band data with a combination of an 
afterglow component described with a broken power-law, a constant 
host galaxy component and a supernova component. We assume that the 
late afterglow, in terms of break-time ($11.5 \pm 3$ day) and post-break decay slope ($2.45 \pm 0.35$), 
follows the values derived from the X-ray afterglow (the pre-break 
decay slope is left free to vary, but is strongly constrained by 
earlier data which is not plotted). The black lines represent the 
sum of all components, directly fit to the data. The blue lines 
give the afterglow plus host galaxy component, whereas the red 
lines show the supernova contribution. Unbroken lines denote the use 
of the central values (M for Mid) derived from the X-ray fit and a 
host galaxy magnitude of $I_C=26.5$. Dotted lines denote the extreme 
(L)ow case of the (within 1$\sigma$ errors of the X-ray fit) earliest 
break and steepest post-break decay slope, as well as the faintest 
host galaxy (where we choose conservative errors of 0.5 magnitudes). 
In this case, at late times, the supernova dominates the afterglow. The dashed lines denote the opposite 
case (H for High), a late break and shallow post-break decay as well 
as a bright host galaxy, which results in the smallest supernova 
contribution, comparable to the afterglow at peak. Even so, we find 
this ``faint'' supernova component to be 50\% more luminous than SN 
1998bw at peak. We also plot the three $z^\prime$ observations. The 
two first points show no color evolution, whereas the last point is 
significantly brighter than the extrapolated $I_C$ light curve and gives further evidence of the strong 
reddening due to the rising supernova component.}
\label{fig:late}
\end{figure*}

\section{Analysis}
\label{sec:als}

\subsection{Optical Light Curve}

We fit the optical data in all filters simultaneously by fitting a
series of summed \cite{Beuermann+1999} functions, a generalization of
the procedure described by \cite{Perley+2008a}.  Several iterations of
different models with varying assumptions and complexity were fitted,
with our final preferred model motivated by a combination of
assumptions of underlying physical behavior (as will be described
shortly) and the necessity to fit the data with a reasonable $\chi^2$
and without large trends in the residuals.  This model contains two elements: a
monochromatic, rapidly falling component which dominates at early
times ($\lesssim 10^4$ s), plus a chromatically evolving second
component which peaks at about $10^3$ -- $10^4$ s (depending on the filter).  
The first
component is actually a sum of three Beuermann functions (this was
necessary to fit several low-level modulation ``wiggles'' in the
data); the second component is a single Beuermann function, but the
peak time is allowed to be a function of the filter central wavelength
to allow the break (peak) to be chromatic.  (Modulations are seen at
late times as well, but we do not have the temporal coverage to
accurately attempt to characterize them.)  We exclude points after 2
$\times$ $10^6$ s, which appear to be contaminated by additional light
(possibly due to a supernova; see \S 3.7) in most filters.  We use
only our own calibrated data; observations reported in the GCN
Circulars are excluded from the fits.  Very early points from UVOT and
PAIRITEL that were heavily affected by pileup, saturation, or
nonlinearity effects were excluded, and a small number of other
conspicuous outliers at late times were manually flagged during the
fitting process.

Our model is found to match the data very well, with no obvious
residual trends, with the exception of a possible chromatic divergence
of the very early-time IR measurements from the initial power law.
The final value for $\chi^2$ = 938 (over 592 degrees of freedom) is
reasonable, if not strictly statistically acceptable, although this is
not necessarily a surprise given the large number of different
instruments and very complicated behavior of the afterglow.  A modest underestimate of the systematic errors could account for $\chi^2/$dof $>1$. All
optical photometry from our measurements and the GCN Circulars is
presented in Figure \ref{fig:lc}. Here we adopt the convention for the spectral index ($\beta$) and the temporal index ($\alpha$) such that $f_\nu \propto \nu^{-\beta}\, t^{-\alpha}$.   A subset of these observations
showing our measurements in more detail is shown in Figure
\ref{fig:lczoom}.

The early afterglow decay is extremely rapid: $\alpha \approx 2.24$.
At about 500~s the light curve flattens slightly, and then levels out
much more significantly around 1500~s, as noted by
\cite{Li08_GCN7438}.  Our rapid temporal sampling ends at 20,000~s due
to bright morning twilight, but we began Gemini observations the
following night, and from then until about $10^6$~s the optical
afterglow decays approximately as a simple power law, although some
limited achromatic variations both above and below the fit may suggest
small modulations.

Our fitting procedure is capable of assigning different colors to
different components and modeling chromatic breaks, so there is no
need to assume achromatic evolution.  Indeed, attempts at
monochromatic fits invariably produced very poor $\chi^2$ values and
obvious residual trends, and color evolution is clear in the
time-dependent SED (\S \ref{sec:sed}).  Our final preferred model is
able to fit all the chromatic evolution using the chromatic break of
the late-time component, which transitions from blue (while rising) to
red (while fading).  This behavior is strongly reminiscent of the
predicted evolution of a forward shock at peak, so we fix the rising
power-law index $\alpha_{1,b}$ and the overall color change over the
break $\Delta\beta_{1,b-a}$ to the values predicted by this model.
(They are nearly unconstrained by the fits otherwise.)  This late-time
component is allowed to have a different post-break color from the
early-time component, but the fits suggest they have similar values,
with the overall beginning-to-end change in spectral index of $\Delta
\beta = 0.006 \pm 0.020$, despite a notable color change in the
intermediate region.

\subsection{SED and Extinction Constraints}
\label{sec:sed}

Due to the excellent coverage of the photometry across the UV/O/IR
spectrum, we can strongly constrain the host-galaxy extinction.  Our
chromatic model can generate SEDs at any point in the evolution of the
light curve using all available filters, shielding us from the
possibility that the intrinsic (pre-extinction) SED may not be a
simple power law during the periods when most of the observations were
actually taken (though only to the extent to which our model
accurately describes the light curve).  We can therefore use all
available filters to fit a single extinction law to the full data in
spite of the observed chromatic behavior.  After correcting for the
small Galactic extinction along this sightline of $E(B - V)$
= 0.01 mag \citep{Schlegel:1998p695} and excluding the Swift UVM2 and UVW2 filters (which are likely to be significantly affected by Lyman-$\alpha$ and Lyman-break absorption, respectively), we measure $A_{V,{\rm host}}$ =
$0.07 \pm 0.06$ mag for a fit to the Small Magellanic Cloud (SMC)
extinction law.  Milky Way Galaxy and Large Magellanic Cloud (LMC)
laws were also tried, but generally had higher values of $\chi^2$ ---
to be sure, the  small amount of extinction makes the exact
choice of extinction law unimportant.  Due to the possibility of
hydrogen absorption and the generally uncertain nature of extinction
laws in the far UV, the UVW2 and UVM2 filters were excluded from 
these fits.  Compared to the X-ray host-galaxy hydrogen column of
$N_{\rm H} = (1.87\pm 0.13) \times 10^{21}$~cm$^{-2}$, this suggests a
dust-to-gas (actually dust-to-free-metals) ratio of about ten times
the Galactic value, which is typical for previous GRB sightlines
\citepeg{Schady+2007}.  The small amount of extinction inferred
is also typical \citepeg{Kann2006}.

We can also generate purely observational time-dependent SEDs.  The
photometric observations taken the night of the burst were divided
into 40 overlapping regions, and a simple power law was fit to the
magnitudes in each region in order to produce contemporaneous observed
fluxes in each available filter at each region.  Galactic extinction,
as well as the small amount of host extinction correction found above,
are removed, and each SED is then fit to a power law to estimate
the overall UV/O/IR spectral index $\beta$ as a function of
time.  Some representative epochs are shown in Figure \ref{fig:seds}.
These are supplemented by SEDs generated directly from the late-time
Gemini observations (with a small correction owing to the fact that
different filters were not observed exactly contemporaneously), shown
in Figure \ref{fig:seds2}.  Finally, we can combine all available data
to plot the spectral index $\beta$ as a function of time, as shown in
Figure \ref{fig:beta}.  The chromatic behavior of this afterglow is
immediately evident.

\subsection{X-ray Fits}

We employ a similar fitting procedure to the X-ray light curve as in
the optical and distinguish two distinct components: a smooth early
decay of $\alpha = 1.48$ breaks at 2500 s to a short lived and poorly
constrained (due to lack of observations during this period) fast
decay of $\alpha \approx 2.8$, and then declines as $\alpha = 1.40$
for the remainder of its evolution (possibly except for the last
observation in which the afterglow is detected, which appears to be
significantly below this curve).  This differs from the optical value
by $\Delta \alpha = 0.17$.  Toward the very end of the XRT
observations, the flux appears to decline more rapidly, and in our
last binned observation is significantly below our fitted model.  This
appears to provide evidence for an additional break at $\sim 2 \times
10^6$ s, but beyond this point the afterglow flux faded below the XRT
detection threshold and further observations are unavailable, so we
cannot independently verify this behavior.

As discussed previously in \S \ref{sec:hep}, the X-ray spectrum
appears invariant over the entire burst evolution within fairly
stringent constraints.  At early times (during the smooth early decay)
we measure $\beta_{X,1}=0.814\pm 0.013$; during the late phase we
measure $\beta_{X,2}=0.80\pm 0.04$.

\subsection{Testing Canonical Models}
\label{sec:canonball}

We are able to measure $\alpha$ and $\beta$ very precisely throughout
most of the evolution of the light curve in both optical and X-ray
bands (limited in most cases only by the assumptions of the model
itself, e.g., the possible presence of additional underlying
components, rather than the actual statistical uncertainty).  In
principle this should allow us a strong test of the canonical fireball
model \citepeg{Sari+1998}.

During the rapid decay phase of the afterglow, we measure (after
extinction correction) $\alpha_{\rm opt}=2.2$, $\beta_{\rm opt}=0.5$ in the
UV-optical-IR range (though both values are somewhat variable).
Unsurprisingly, this combination of large $\alpha$ and low $\beta$ is
not well-fit by any canonical forward-shock model
(e.g., \citealt{Price:2003p575}), with the exception of a jet model
where $\nu_{\rm opt} < \nu_c$, which is approximately consistent with
both values for $p = 2.0-2.2$.  We will return to the possibility of
an extremely early jet break in \S \ref{sec:energy}, but the later
evolution of the light curve (in which it flattens) casts doubt on
this interpretation.  Alternatively, the reverse-shock model
\citep{Kobayashi2000} also predicts a very steep early decay, although
even this model cannot exactly reconcile the steep decay with the
observed spectral index; given $\alpha = 2.2$ this model predicts
$\beta$=0.8.  Still, this is a significant improvement over any
forward-shock model.  A third possibility, favored by
\cite{Kumar+2008}, is that the early steep decay is actually ``high
latitude'' emission: prompt emission whose arrival at the observer was
delayed because it is slightly off-axis.  In this case a value of
$\beta = \alpha - 2$ = 0.2 is predicted, which is also not consistent
with our observations.  In fact, the observed value is nearly exactly
between the reverse-shock and high-latitude values.  It is possible that both
models may contribute in about equal degrees to the observations and
produce an intermediate spectral index (which may also explain some of
the small-scale variation in the observed $\alpha$).

At about the same time when the optical light curve is rapidly
decaying, the X-ray light curve is relatively flat, with $\alpha_X$ =
1.48.  Comparing the relative optical and X-ray fluxes, the optical
flux overpredicts the X-ray flux at early times; the optical and
X-ray bands also have different spectral indices, suggesting that a
spectral break is present between the two bands and that this break is
moving upward in frequency with time.  Given the disparate behavior
from the contemporaneous optical light curve and the shift to different
behavior at late time, the early X-ray emission appears to come from
a different origin or process as the rest of the light curve; we do
not attempt to explain it within the standard model.

Once the optical light curve begins to level out (in the transition
zone between the early and late-time components), the spectral index
evolves to surprisingly blue values.  An average SED formed from data
points during the extremely well-observed region, using observations
from $K$ through the UV between 5500 and 12,000~s, constrains the
average SED to $\beta_{\rm opt} = 0.07 \pm 0.03$ (see also Figure
\ref{fig:beta}).  It is impossible to generate $\beta < 0.5$ during a
declining light curve within the basic synchrotron model, so this
offers strong corroborating evidence that the underlying component
responsible for the majority of the optical flux is undergoing a break
at this time.  The  canonical model \citep{Sari+1998} could
naturally explain this portion of the light curve as a
characteristic-frequency peak, when $\nu_m$ crosses through the UV,
optical, and IR bands (in that order), causing a gradual shift in the
overall color from $\beta$ = $-1/3$ to $\beta$ = $(p-1)/2$.  It is in
fact difficult to explain this feature in any other straightforward
way; in a wind medium the same feature shifts in color only from
$\beta$ = $1/2$ to $\beta$ = $p/2$ \citep{LiChevalier2003} which is
never sufficiently blue.  Complicated prescriptions involving numerous
components contributing about equally to the afterglow flux
\citepeg{Racusineteveryone} may also explain this feature, though at
the expense of substantial additional complexity.

Although the optical SED provides good evidence of shifting somewhat
back toward the red at late times, its overall blueness remains a
challenge to explain even after a simple power-law decay has set in.
This late decay rate is $\alpha_{\rm opt} = 1.23 \pm 0.02$ optically,
and $\alpha_X = 1.30 \pm 0.05$ in X-rays.
Fitting to a combination of the first three epochs of the late-time
Gemini data gives $\beta_{\rm opt} = 0.51 \pm 0.26$.  If we assume
that the X-ray and optical afterglows have the same origin, then the
X-ray to optical spectral index is $\beta_{\rm OX} = 0.77 \pm 0.01$ ---
consistent with the X-ray index itself of $\beta_{\rm X}=0.80\pm 0.04$.
This also forms a strict upper limit on $\beta_{\rm opt}$; even if a
cooling break is present close to the optical band (our observations
do not rule out this possibility) then $\beta_{\rm opt} < \beta_{\rm OX}$
to not underpredict the X-ray flux.

Given this, and assuming $p \geq 2$, we can greatly constrain the
available models.  The jet model (which we favored in the first
version of this paper, before late-time data were available) is
completely ruled out (it requires $\alpha \geq 2.0$), as are models
where the optical band is above the cooling break (which requires
$\beta \geq 1.0$).  This leaves only two possibilities.  The
interstellar medium (ISM) model discussed previously is, in principle,
still in agreement (barely) with the data: for $p$ = 2.59, $\alpha$ =
1.19 and $\beta$ = 0.79, all within 2$\sigma$ of observed values.
However this requires that the X-ray regime also be below the cooling frequency
($\nu_X < \nu_c$) out to extremely late times (at least 10~d),
which is not expected physically in the ISM model unless the density
$n$ and/or the magnetization fraction $\epsilon_B$ are very low:
\begin{equation}
\left(\frac{n}{\rm cm^{-3}}\right)\left(\frac{\epsilon_B}{\rm 4 \times
10^{-5}}\right)^{3/2} \lesssim 1. \nonumber
\end{equation}  
\noindent Alternatively, a wind model with $p =
2.0$ predicts $\alpha_{\rm opt} = 1.25$ and $\beta_{\rm opt}$ = 0.50.
These are reasonable, but the X-ray flux is either greatly
over-predicted (if a cooling break is absent, we would anticipate
$\beta_{\rm OX}$ = 0.50 as well), or the X-ray decay is far too fast (if a
cooling break is present, then $\alpha_{\rm X}$ = 1.0).  Either
possibility is ruled out to at least 6$\sigma$.

We therefore prefer the ISM model to favor all observations.  While a
wind-stratified medium may be made to work with sufficient
enhancements (for example, the chromatic optical behavior could be
explained by an additional emission component, and the late-time
discrepancies in the optical and X-rays could be explained by evolution
of the cooling frequency in a way not predicted by canonical models),
this model is simpler as it explains all of our observations without
the need for such components --- with the exception of the very
early-time data, where standard assumptions are most expected not to hold.  Our
conclusion here therefore disagrees with that of \cite{Kumar+2008}
and \cite{Racusineteveryone}, who prefer a wind-stratified model at
late times.

\subsection{Frequency Domain Searches}

Given the brightness of the event, we analyzed the high time cadence
$\gamma$-ray, X-ray, and IR data in search of any significant trends
in the frequency domain. The power-density spectrum (PDS) for the {\it
INTEGRAL} data \citep{beckmann2008}, spanning $f = 0.03$ to $5$\,Hz,
is well fit by a power law with $P(f) df \propto f^{-1.8}$.  This
index here is consistent with the $-5/3$ reported for a number of
bright GRBs \citep{bss98,cy00}, and is interpreted as evidence for fully
developed turbulence.  The PDS from the {\swift} XRT data is
essentially (white) noise, apart from $1/f$ noise at short
frequencies. An analysis of the first $\sim$2 hr of PAIRITEL
$J$-band photometry also shows a featureless PDS aside from $1/f$
noise.

\begin{figure*}[p] 
\centerline{\psfig{file=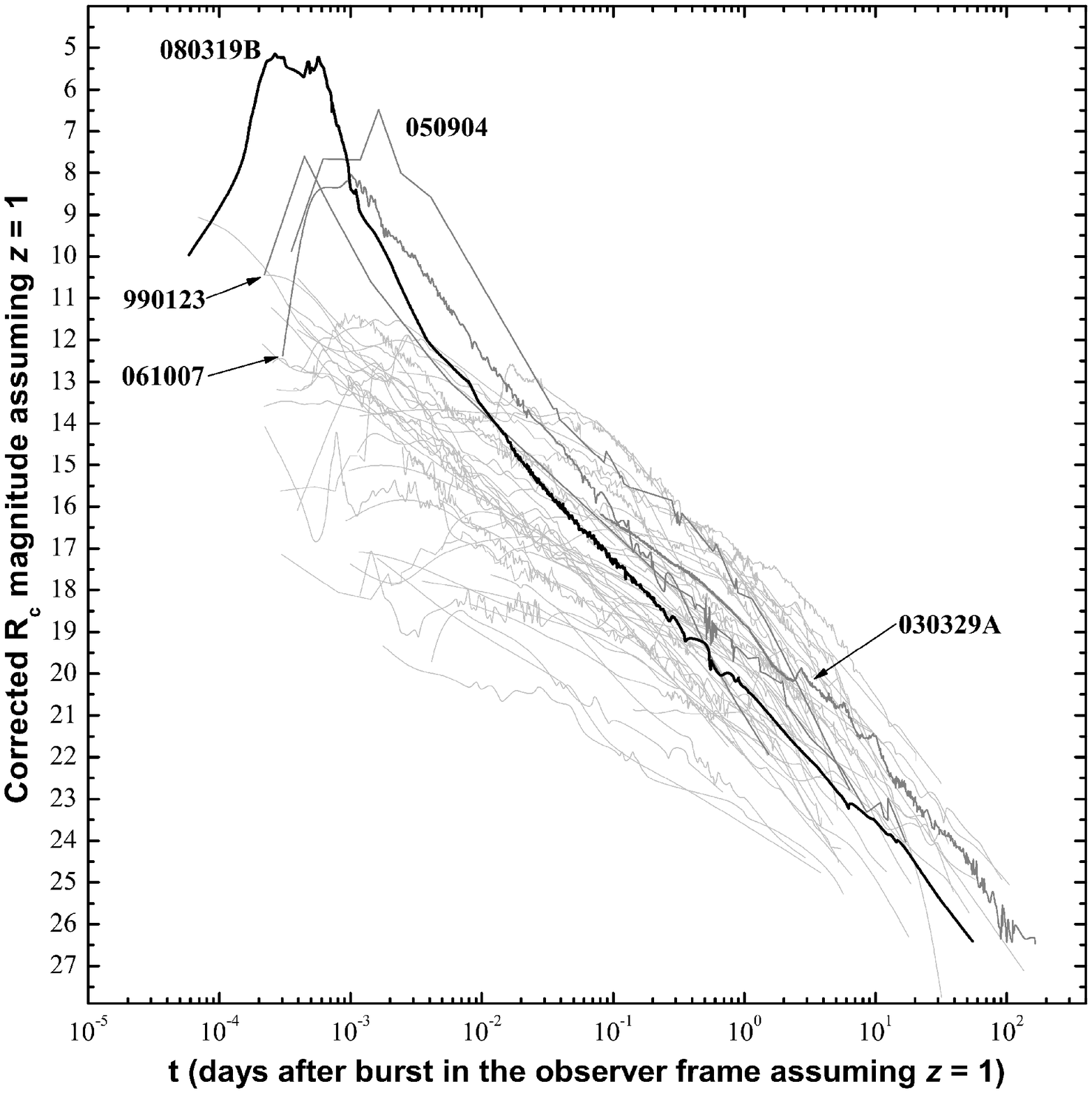,width=5.5in,angle=0}}
\caption[] {\footnotesize Comparison between the observed $R$-band light curve of
\thegrb and those of other GRB afterglows, both from the pre-\swift as
well as the \swift era, shifted to a common redshift of $z=1$ with the
method of \cite{Kann2006}.  The prompt flash of \thegrb is clearly
shown to be the most luminous optical transient ever observed with a
high degree of confidence.  In spite of this, because of its rapid
early decay the afterglow at late times is quite unremarkable, and is
similar in this regard to the three other ``ultra-luminous''
bursts to date: GRBs~990123, 061007, and 050904.  In contrast,
the bursts that remain the brightest tend to be those with late
plateaus and slow decays.}
\label{fig:z1}
\end{figure*}

\subsection{Energetics}
\label{sec:energy}

The emission from GRB~080319B makes it one of the most energetic GRBs
ever detected and the most energetic thus far seen by {\swift}.  A
preliminary analysis of Konus-Wind observations reported by
\citet{gam+08} yields a burst fluence of ${5.72}_{-0.13}^{+0.14}$
$\times 10^{-4}$~erg cm$^{-2}$ in a 20 keV to 7 MeV energy window.
This corresponds to a fluence of $5.31 \times 10^{-4}$~erg cm$^{-2}$
in the BATSE bandpass of 20~keV to 1.8~MeV, making it brighter than
the highest-fluence BATSE burst \citep[$4.08 \times 10^{-4}$~erg
cm$^{-2}$;][]{Kaneko06}.  The rest-frame isotropic equivalent energy
release is $E_{\rm iso}= 1.3 \times 10^{54}$~erg in the standard
source frame $1-10^4$~keV band \citep{Amati02,Bloom01}.

If the true collimation-corrected energy ($E_{\gamma}$) of \thegrb is
comparable to the median value of $E_{\gamma} = 1.33 \times
10^{51}$~erg found by \citet{Bloom03}, and if we assume average values
for the efficiency of converting the blast wave's kinetic energy into
gamma rays of $\eta_{\gamma}=0.5$ and the circumburst density
$n=1.0$~cm$^{-3}$ \citep{Granot06,Kumar07}, then a jet break is
expected at $3.6 \times 10^{4}$~s post trigger.  A circumburst density
greater than $n=1$~cm$^{-3}$ only hastens the predicted light-curve
break, with $n=100$~cm$^{-3}$ resulting in $t_{\rm jet} =7.8 \times
10^3$~s.  Likewise, a collimation-corrected energy less than
$E_{\gamma} = 1.3 \times 10^{51}$~erg will also push the expected jet
break to earlier times, with $E_{\gamma} = 10^{50}$~erg yielding
$t_{\rm jet} \approx 10^{3}$~s.

No canonical jet break is observed in the GRB light curve.  The only
unambiguous steepening feature is an X-ray break observed at $t = 2620
\pm 470$~s.  However, no contemporaneous optical break is observed.
There is evidence of a break at very late times ($\gtrsim 10^6$~s).
Unfortunately, the afterglow is extremely faint at this point, but
both the last X-ray observation, the last $g$-band detection, and the
final {\it HST} $F606W$ measurement from the GCN circulars are
significantly (though only mildly: 2--3$\sigma$) below our fitted
power-law extrapolation.  Furthermore, the fact that the afterglow
shifts to dramatically redder colors at this point (simultaneously,
$i$ and $z$ rise above the fitted curve) suggests that the optical
flux may be dominated by supernova light (\S 3.7; host-galaxy light likely
contributes as well), and the afterglow contribution is actually even
smaller than this, which would strongly favor a break.  Unfortunately,
given the lack of available late-time X-ray data we cannot strongly
constrain whether or not the break is chromatic.

Treating this feature as a jet break, the very late break time ($t =
10$ d) imposes strong demands on the afterglow energetics.  Assuming
$n=1$~cm$^{-3}$ and $\eta_{\gamma} = 0.5$, this places a lower limit
on the collimation-corrected energy of $4.0 \times 10^{51}$~erg,
significantly higher than any pre-\swift value \citep{Bloom03} and the
vast majority of \swift bursts, but comparable to the handful of
``hyper-energetic'' outlier events discovered by \swift: GRBs 050904
\citep{Tagliaferri+2005}, 050820A \citep{Cenko+2006}, and 070125
\citep{Chandra+2008,Updike+2008}.  The lack of an early jet break would then argue
that the extreme brightness of this GRB is at least partially
intrinsic to the explosion itself.

Alternatively, a jet break may have occurred extremely early (within
the first 100~s) before the start of our observations.  Such a model
has been suggested to explain the lack of an apparent jet break in
another powerful event detected by {\swift}, such as GRB~061007 ---
see \citealt{Schady07} and \citealt{Mundell07}). In this case,
$E_{\gamma}$ would actually have an unusually \emph{low}, though not
unprecedented, energy release of $\sim 10^{49}$~erg.  One could also
appeal to the minor break in the X-ray light curve (and presume the
optical break was hidden by other effects), which would indicate a
fairly canonical energy near $10^{51}$ erg; either of these cases
would point to extreme collimation given the large $E_{\rm iso}$.
However, as we have already discussed, in these cases the
``post-break'' light-curve slopes in the optical and X-rays of
$\alpha_{\rm O} = 1.23$ and $\alpha_{\rm X} = 1.40$ are much too
shallow for the prediction of $t^{-p}$ \citep{Sari1999}, where
generally we expect the electron index $p \geq 2.0$.

A final possibility is that multiple jets, with very different opening
angles, were involved, similar to the model proposed for GRB 030329
\citep{Berger+2003}.  In this scenario, we might expect to see both an
early break (from highly collimated, highly relativistic emission)
\emph{and} a late break (from less collimated, less relativistic
emission), presuming that the emission at different times was
dominated by different jet components.  Assuming that the prompt
emission and optical flash were associated with the narrow jet, the
lack of a break at late times\footnote{To be sure, \citet{Racusineteveryone} claim a late-time jet break but we find no significant evidence of such in the XRT data alone. The apparent discrepancy may be due to the different time binnings of the late-time data.} need not imply unusual energetics.  
The fact that the early-time X-ray break is very sharp, and also
apparently achromatic, might argue for this interpretation, and the
lack of a contemporaneous optical break could be explained if the
optical counterpart of the early X-ray shallow decay phase were hidden
by the bright reverse shock.  This model is the one favored by
\cite{Racusineteveryone}.

\subsection{Supernova Constraints}

Detailing the nature of the late-time behavior and the evolution of
the likely supernova counterpart is beyond the scope of this paper,
but we performed a few basic fits using the method of \cite{Zeh+2004}
of the late-times observations under various model assumptions (host
brightnesses, using an extrapolation of the late-time light curve and
a SN 1998bw-like supernova template).  In all assumptions we measure a luminosity relative to SN
1998bw in the same rest-frame epoch, of greater than one, with an
overall distribution of $k$ = $2.3 \pm 0.8$ (Figure \ref{fig:late}).  In particular, the
existence of a late afterglow break and host galaxy with a magnitude
comparable to the HST measurement of \citet{Levan08_GCN7710} favor higher values,
which provides evidence that this unusually bright GRB may be
associated with an unusually bright (even compared to other GRB
supernovae: \citealt{Ferrero+2006}) supernova.  For $s$, the relative
rest-frame time to maximum light relative to SN 1998bw, we measure 
$s$ = $0.89 \pm 0.10$.

\section{Discussion and Conclusions}

The PDS analysis reflects quantitatively what is qualitatively a light
curve without strong brightening features. Such ``smooth'' afterglows
have been seen before (e.g., \citealt{Laursen:2003p4600, Stanek2005}),
yet many GRBs with high-quality optical/IR afterglow observations show
significant jaggedness, often on timescales less than the time since
trigger \citepeg{Jakobsson:2004p4601}.  The afterglow in this case is
not entirely featureless; slow modulations of less than 10\% with
$\Delta t / t$ are observed at around 200~s and 700~s, but these
have been seen before in other afterglows, including the undulating
afterglow of GRB 030329 \citep{Lipkin:2004p552}.

While an observation of periodicity ($f \age 1$~Hz) in the prompt
phase could be considered a reasonable manifestation of a magnetized
neutron star powering the initial internal shocks
\citepeg{Metzger:2007p4684}, the absence of such a signature in the
{\it INTEGRAL} PDS by no means rules out such a hypothesis.  The
concordance of the PDS with other GRBs in the pre-{\swift} sample
suggests, {\it prima facie}, a similarity between \thegrb and other
events in the hydrodynamic properties of the (emitting) outflow; also
absent is any substantial evidence for a difference in microphysical
parameters. We are left with the reasonable conclusion that the
extreme brightness of \thegrb has more to do with macroscopic
parameters of the central engine (in particular, the collimation
angle, $M_{\rm ejecta}$, initial Lorentz factor, and possibly the
circumburst medium could all be responsible) than extrema in shock
parameters.

\begin{figure*}[p]
\centerline{\psfig{file=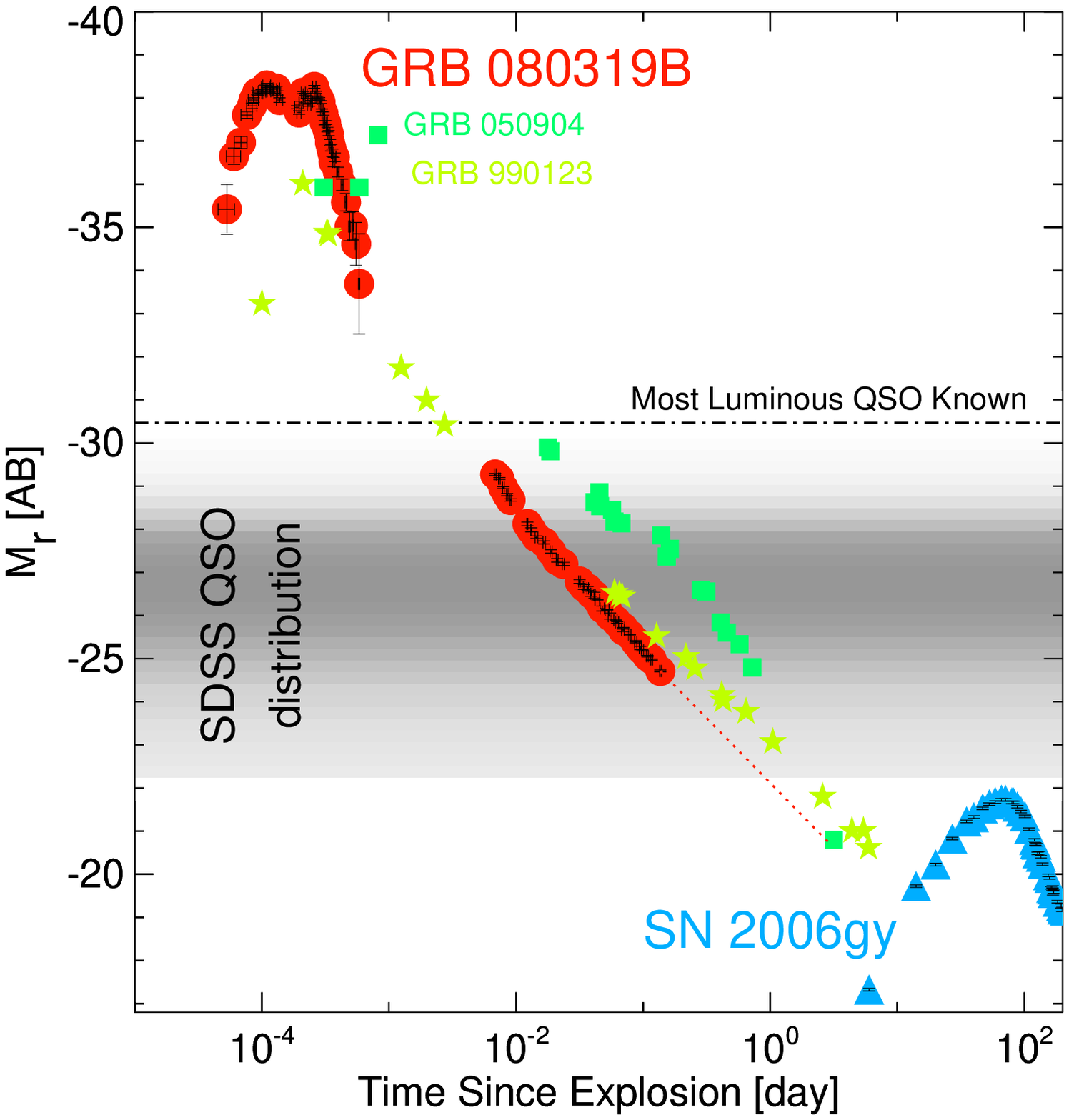,width=5.0in,angle=0}}
\caption[] {\small Rest-frame comparison of the most luminous
optical/IR probes of the distant universe, showing the absolute
magnitude ($M_r$, in AB magnitudes as defined by \citealt{oke82})
versus time of \thegrb (red circles) and SN\,2006gy (blue triangles;
\citealt{Smith:2007p532}). Transformed light curves of GRB\,990123
(yellow stars; adapted from \citealt{gbw+99}) and GRB\,050904 (green
squares; adapted from \citealt{Kann:2007p133}) are also shown. For
reference, the most luminous known QSO \citep{Schneider:2007p4602} is
shown with a dashed horizontal line; the distribution of SDSS QSO
magnitudes, adapted from Fig.\ 6 of \citeauthor{Schneider:2007p4602},
is shown as horizontal banding (darker indicates higher density of
sources per unit magnitude). The afterglow of \thegrb was the
brightest GRB afterglow ever recorded and was at early times
$\sim$10$^{3}$ times more luminous than the most luminous QSO.}
\label{fig:vsqso}
\end{figure*}

\begin{figure*}[p]
\centerline{\psfig{file=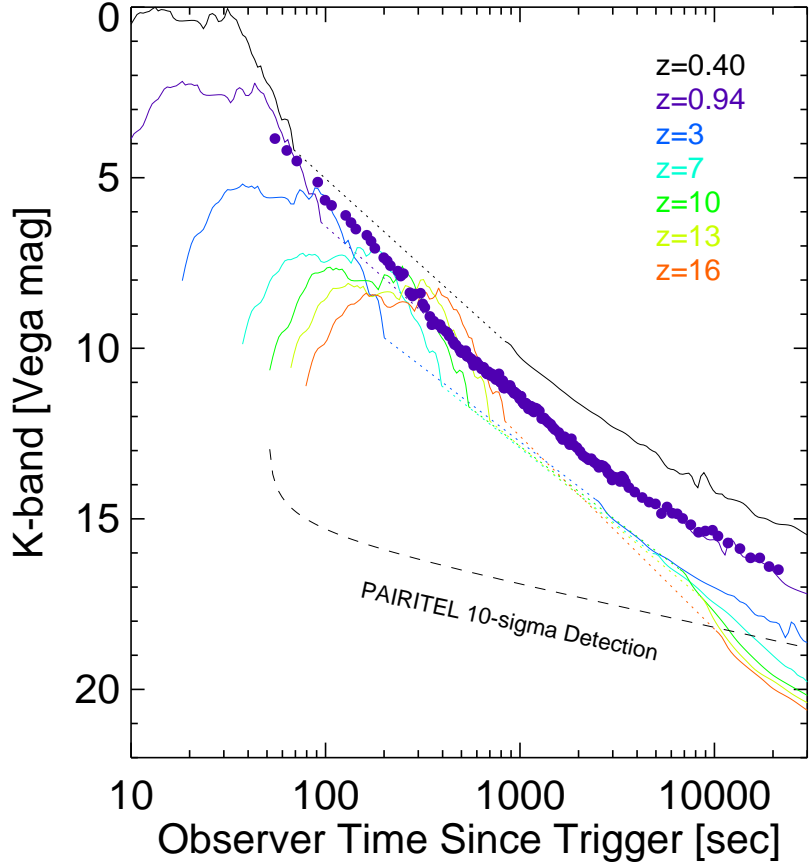,width=5.0in,angle=0}}
\caption[] {\small 
Observability of \thegrb in observer-frame $K$ band at a variety of
redshifts, based on a simple model of the temporal and spectral
evolution of the afterglow bootstrapped from the $V$-band light curve.
Dotted lines cover the gap in optical ($V$-band) coverage.  The
afterglow, which would have been as bright as Vega if the event
occurred at $z=0.40$, remains remarkably bright even to $z \approx
16$. The purple filled circles show the observed $K$-band light curve
from this paper, showing a good agreement with the model of the
$z = 0.937$ (purple) curve (suggesting, too, that the interpolation in the V-band time gap is appropriate). The PAIRITEL cumulative 10$\sigma$
point-source sensitivity is shown, assuming a nominal start time after
GRB trigger of 51~s. For all redshifts where the Universe is
transparent to Ly$\alpha$ photons, this source could have easily been
detected by sub-meter class telescopes.}
\label{fig:z15}
\end{figure*}

There is a qualitative similarity in the intrinsic behavior of the
three brightest afterglow events (Fig.\ \ref{fig:z1}), requiring more
than 9~s in the rest frame from $\gamma$-ray trigger to reach peak
brightness. Yet, as noted with GRBs 050904 and 990123
\citep{Kann:2006p3171}, even the brightest events fade very rapidly
and reach by day 1 a magnitude comparable to that of the general
population. In our adopted cosmology, $M_{{\rm peak},r} = -38.3$ mag
at a time $t \approx 10$ s (rest frame) from trigger.  Likewise, $M_{{\rm
peak},r} (990123) = -36.0$ mag at a time $t=18.1$ s and $M_{{\rm
peak},r} (050904) = -37.1$ mag at a time $t=71.2$ s. Compared to GRBs
990123 and 050904, \thegrb does not appear to support the proposed
brightness vs. $t_{\rm peak}$ relation \citep{Panaitescu:2008p3168}
(i.e., that brighter events are seen to take longer from trigger to
peak); still, the existence of a rough relation when including fainter events suggests that the very
brightest events such as these may allow for the longest follow-up delay at a given
redshift. This afterglow ``turn on'' delay for some fraction of events has
important implications for follow-up of high-redshift GRBs since tens
of seconds in the rest frame becomes a delay of minutes in the
observer frame (Fig.\ \ref{fig:z15}). Such a consideration, for
example, could relax requirements for extremely rapid repointing of
satellites to catch afterglows at their brightest.

Figure \ref{fig:vsqso} shows the comparison between GRB\,080319B,
quasars, and one of the most energetic supernova recorded (SN~2006gy;
\citealt{Smith:2007p532}). While evolutionary effects in all three
populations are sure to be important at some level (with QSOs fainter
at higher redshift, etc.), in the context of probing the high-redshift
universe, the overall impression is clear: for $\sim$30 min in the
rest frame (what would be $\sim4$~hr in the observer frame at $z=7$), GRB
080319b would have been brighter than the brightest known QSO in the universe (see also \citealt{2001grba.conf..191L,Kann2006} for a discussion comparison of the GRB population with QSOs).

We now turn our attention to {\em detectability} of such GRBs at high
redshift (leaving aside the question of their {\em existence}; see,
e.g., \citealt{BrommLoebReview,Naoz+2007}).  Given the observed light
curve and intrinsic spectrum of this burst ($\S$ 2), to what redshift
could similar events be detected by present and future missions?  The
peak photon flux from GRB~080319B in 1~s is $\sim 1 \times
10^{59}$~photons s$^{-1}$.  For a BAT threshold flux of 0.8 photons
cm$^{-2}$ s$^{-1}$ for $E_{\rm pk,obs} \gtrsim 100$~keV
\citep{band03}, the event would be detected out to $z=10.7$
($8\sigma$).  For the nominal EXIST \citep{2007AIPC..921..211G}
threshold of $0.2$~photons~cm$^{-2}$ s$^{-1}$, the event would be
detected out to $z=32$.  The ability to detect an afterglow for an
extreme event of this nature is even more remarkable because of
time-dilation effects --- shifted to $z \approx 32$ or potentially
even further, the afterglow luminosity at late times is nearly
independent of distance at high redshift.  Aside from the effects of
line-of-sight absorption by neutral hydrodgen, \thegrb would remain visible even if placed well into the
epoch of reionization (Fig.\ \ref{fig:z15}).

We conclude with a rumination on the extrapolation of the features of
the afterglow toward the low-redshift universe.  At $z=0.17$, the
distance of the nearest non-underluminous GRB to date (GRB
030329), this event would peak at $R \approx 1$ mag, nearly as bright as
the brightest stars in the sky.  Or, to carry the comparison to its
greatest extreme, we might envision a situation in which a GRB similar
to \thegrb were to occur in our own Galaxy.  At a distance of 1 kpc
(and neglecting the probable substantial extinction along Galactic
lines of sight at optical wavelengths), the optical flash would peak
at magnitude about $-28.5$, several times the brightness of the Sun!
Such an event must assuredly be extremely unusual --- the Galactic GRB
rate is probably no greater than 1 per $10^5$ to $10^6$ years
(\citealt{Posiadlowski2004}; see also \citealt{Stanek2006}), likely
only 1\% of such bursts are collimated toward Earth, and this is
among the brightest 0.1\% of bursts ever observed.  Altogether, the
rate is probably less than 1 per $10^{10}-10^{11}$ yr: unlikely to
have ever happened even over the long timescale of geological history,
and certainly not a spectacle we can expect to witness anytime soon.

\acknowledgements

Much of this research could not have been undertaken without the work
of S. Barthelmy running the GCN Circulars and the {\swift} team for
their extraordinary efforts.  N.R.B. was partially supported by a
SciDAC grant from the Department of Energy. J.S.B., J.X.P., and
H.-W.C. are partially supported by NASA/{\swift} grant \#NNG05GF55G.
A.V.F.'s group at U.C. Berkeley is supported by National Science
Foundation (NSF) grant AST--0607485 and the TABASGO Foundation, as
well as by NASA/{\swift} Guest Investigator grants \#NNG05GF35G and
\#NNG06GI86G.  S.L. was partly supported by the Chilean {\sl Centro de Astrof\'\i sica}
FONDAP No. 15010003 and by FONDECYT grant N$^{\rm o}1060823$. We thank J.\ Brewer of the IFMD Foundation. This
publication makes use of data products from the Two Micron All Sky
Survey, which is a joint project of the University of Massachusetts
and the Infrared Processing and Analysis Center/California Institute
of Technology, funded by NASA and the NSF.  KAIT and its ongoing
research were made possible by donations from Sun Microsystems, Inc.,
the Hewlett-Packard Company, AutoScope Corporation, Lick Observatory,
the NSF, the University of California, the Sylvia \& Jim Katzman
Foundation, and the TABASGO Foundation. Partially, based on observations obtained at the Gemini Observatory, which is operated by the
Association of Universities for Research in Astronomy, Inc., under a cooperative agreement
with the NSF on behalf of the Gemini partnership: the National Science Foundation (United
States), the Science and Technology Facilities Council (United Kingdom), the
National Research Council (Canada), CONICYT (Chile), the Australian Research Council
(Australia), Ministério da Ciência e Tecnologia (Brazil) and SECYT (Argentina).  We thank the
Gemini Observatory staff for their excellent work in observing
GRB\,080319B; the assistance of the Lick Observatory staff is also
acknowledged.  The Peters Automated Infrared Imaging Telescope is operated by the Smithsonian Astrophysical Observatory (SAO) and was made possible by a grant from the Harvard University Milton Fund, the camera loan from the University of Virginia, and the continued support of the SAO and UC Berkeley. The PAIRITEL project is further supported by NASA/Swift Guest Investigator Grant \#NNG06GH50G. We are grateful to E.\ E.\ Falco and the Mt.\ Hopkins staff (W.~Peters, R.\ Hutchins, T.\ Groner) for their continued assistance with PAIRITEL.


\begin{thebibliography}{91}
\expandafter\ifx\csname natexlab\endcsname\relax\def\natexlab#1{#1}\fi

\bibitem[{{Adelman-McCarthy} {et~al.}(2008){Adelman-McCarthy}, {Ag{\"u}eros},
  {Allam}, {Allende Prieto}, {Anderson}, {Anderson}, {Annis}, {Bahcall},
  {Bailer-Jones}, {Baldry}, {Barentine}, {Bassett}, {Becker}, {Beers}, {Bell},
  {Berlind}, {Bernardi}, {Blanton}, {Bochanski}, {Boroski}, {Brinchmann},
  {Brinkmann}, {Brunner}, {Budav{\'a}ri}, {Carliles}, {Carr}, {Castander},
  {Cinabro}, {Cool}, {Covey}, {Csabai}, {Cunha}, {Davenport}, {Dilday}, {Doi},
  {Eisenstein}, {Evans}, {Fan}, {Finkbeiner}, {Friedman}, {Frieman},
  {Fukugita}, {G{\"a}nsicke}, {Gates}, {Gillespie}, {Glazebrook}, {Gray},
  {Grebel}, {Gunn}, {Gurbani}, {Hall}, {Harding}, {Harvanek}, {Hawley},
  {Hayes}, {Heckman}, {Hendry}, {Hindsley}, {Hirata}, {Hogan}, {Hogg}, {Hyde},
  {Ichikawa}, {Ivezi{\'c}}, {Jester}, {Johnson}, {Jorgensen}, {Juri{\'c}},
  {Kent}, {Kessler}, {Kleinman}, {Knapp}, {Kron}, {Krzesinski}, {Kuropatkin},
  {Lamb}, {Lampeitl}, {Lebedeva}, {Lee}, {Leger}, {L{\'e}pine}, {Lima}, {Lin},
  {Long}, {Loomis}, {Loveday}, {Lupton}, {Malanushenko}, {Malanushenko},
  {Mandelbaum}, {Margon}, {Marriner}, {Mart{\'{\i}}nez-Delgado}, {Matsubara},
  {McGehee}, {McKay}, {Meiksin}, {Morrison}, {Munn}, {Nakajima}, {Neilsen},
  {Newberg}, {Nichol}, {Nicinski}, {Nieto-Santisteban}, {Nitta}, {Okamura},
  {Owen}, {Oyaizu}, {Padmanabhan}, {Pan}, {Park}, {Peoples}, {Pier}, {Pope},
  {Purger}, {Raddick}, {Re Fiorentin}, {Richards}, {Richmond}, {Riess}, {Rix},
  {Rockosi}, {Sako}, {Schlegel}, {Schneider}, {Schreiber}, {Schwope}, {Seljak},
  {Sesar}, {Sheldon}, {Shimasaku}, {Sivarani}, {Smith}, {Snedden}, {Steinmetz},
  {Strauss}, {SubbaRao}, {Suto}, {Szalay}, {Szapudi}, {Szkody}, {Tegmark},
  {Thakar}, {Tremonti}, {Tucker}, {Uomoto}, {Vanden Berk}, {Vandenberg},
  {Vidrih}, {Vogeley}, {Voges}, {Vogt}, {Wadadekar}, {Weinberg}, {West},
  {White}, {Wilhite}, {Yanny}, {Yocum}, {York}, {Zehavi}, \& {Zucker}}]{DR6}
{Adelman-McCarthy}, J.~K. {\it et al.} 2008, \apjs, 175, 297

\bibitem[{{Amati} {et~al.}(2002){Amati}, {Frontera}, {Tavani}, {in't Zand},
  {Antonelli}, {Costa}, {Feroci}, {Guidorzi}, {Heise}, {Masetti}, {Montanari},
  {Nicastro}, {Palazzi}, {Pian}, {Piro}, \& {Soffitta}}]{Amati02}
{Amati}, L. {\it et al.} 2002, \aap, 390, 81

\bibitem[{{Band}(2003)}]{band03}
{Band}, D.~L. 2003, \apj, 588, 945

\bibitem[{{Beckmann} {et~al.}(2008){Beckmann}, {Mereghetti}, {Kienlin}, {Beck},
  {Savchenko}, {Borkowski}, \& {Gotz}}]{beckmann2008}
{Beckmann}, V., {Mereghetti}, S., {Kienlin}, A.~v., {Beck}, M., {Savchenko},
  V., {Borkowski}, J., and {Gotz}, D. 2008, {GCN Circular} 7450

\bibitem[{{Beloborodov} {et~al.}(1998){Beloborodov}, {Stern}, \&
  {Svensson}}]{bss98}
{Beloborodov}, A.~M., {Stern}, B.~E., and {Svensson}, R. 1998, \apjl, 508, L25

\bibitem[{{Berger} {et~al.}(2003){Berger}, {Kulkarni}, {Pooley}, {Frail},
  {McIntyre}, {Wark}, {Sari}, {Soderberg}, {Fox}, {Yost}, \&
  {Price}}]{Berger+2003}
{Berger}, E. {\it et al.} 2003, \nat, 426, 154

\bibitem[{{Bertin} \& {Arnouts}(1996)}]{SExtractor}
{Bertin}, E. and {Arnouts}, S. 1996, \aaps, 117, 393

\bibitem[{{Beuermann} {et~al.}(1999){Beuermann}, {Hessman}, {Reinsch},
  {Nicklas}, {Vreeswijk}, {Galama}, {Rol}, {van Paradijs}, {Kouveliotou},
  {Frontera}, {Masetti}, {Palazzi}, \& {Pian}}]{Beuermann+1999}
{Beuermann}, K. {\it et al.} 1999, \aap, 352, L26

\bibitem[{{Bloom} {et~al.}(2003){Bloom}, {Frail}, \& {Kulkarni}}]{Bloom03}
{Bloom}, J.~S., {Frail}, D.~A., and {Kulkarni}, S.~R. 2003, \apj, 594, 674

\bibitem[{{Bloom} {et~al.}(2001){Bloom}, {Frail}, \& {Sari}}]{Bloom01}
{Bloom}, J.~S., {Frail}, D.~A., and {Sari}, R. 2001, \aj, 121, 2879

\bibitem[{{Bloom} {et~al.}(2008){Bloom}, {Starr}, \& {Perley}}]{bsp08}
{Bloom}, J.~S., {Starr}, D., and {Perley}, D.~A. 2008, {GCN Circular} 7434

\bibitem[{{Bloom} {et~al.}(2006){Bloom}, {Starr}, {Blake}, {Skrutskie}, \&
  {Falco}}]{Bloom:2006p4665}
{Bloom}, J.~S., {Starr}, D.~L., {Blake}, C.~H., {Skrutskie}, M.~F., and
  {Falco}, E.~E. 2006, in Astronomical Society of the Pacific Conference
  Series, Vol. 351, Astronomical Data Analysis Software and Systems XV, ed.
  C.~{Gabriel}, C.~{Arviset}, D.~{Ponz}, \& S.~{Enrique}, 751

\bibitem[{{Bromm} \& {Loeb}(2007)}]{BrommLoebReview}
{Bromm}, V. and {Loeb}, A. 2007, in American Institute of Physics Conference
  Series, Vol. 937, Supernova 1987A: 20 Years After: Supernovae and Gamma-Ray
  Bursters, ed. S.~{Immler}, K.~{Weiler}, \& R.~{McCray}, 532--541

\bibitem[{{Butler}(2008)}]{but08}
{Butler}, N. 2008, {GCN Circular} 7499

\bibitem[{{Butler} \& {Kocevski}(2007{\natexlab{a}})}]{Butler:2007p4519}
{Butler}, N.~R. and {Kocevski}, D. 2007{\natexlab{a}}, \apj, 663, 407

\bibitem[{{Butler} \& {Kocevski}(2007{\natexlab{b}})}]{bnk07b}
---. 2007{\natexlab{b}}, \apj, 668, 400

\bibitem[{{Butler} {et~al.}(2007){Butler}, {Kocevski}, {Bloom}, \&
  {Curtis}}]{Butler:2007p4518}
{Butler}, N.~R., {Kocevski}, D., {Bloom}, J.~S., and {Curtis}, J.~L. 2007,
  \apj, 671, 656

\bibitem[{{Cenko} {et~al.}(2006){Cenko}, {Kasliwal}, {Harrison}, {Pal'shin},
  {Frail}, {Cameron}, {Berger}, {Fox}, {Gal-Yam}, {Kulkarni}, {Moon}, {Nakar},
  {Ofek}, {Penprase}, {Price}, {Sari}, {Schmidt}, {Soderberg}, {Aptekar},
  {Frederiks}, {Golenetskii}, {Burrows}, {Chevalier}, {Gehrels}, {McCarthy},
  {Nousek}, \& {Piran}}]{Cenko+2006}
{Cenko}, S.~B. {\it et al.} 2006, \apj, 652, 490

\bibitem[{{Chandra} {et~al.}(2008){Chandra}, {Cenko}, {Frail}, {Chevalier},
  {Macquart}, {Kulkarni}, {Bock}, {Bertoldi}, {Kasliwal}, {Fox}, {Price},
  {Berger}, {Soderberg}, {Harrison}, {Gal-Yam}, {Ofek}, {Rau}, {Schmidt},
  {Cameron}, {Cowie}, {Cowie}, {Dopita}, {Peterson}, \&
  {Penprase}}]{Chandra+2008}
{Chandra}, P. {\it et al.} 2008, ArXiv e-prints, 0802.2748

\bibitem[{{Chang} \& {Yi}(2000)}]{cy00}
{Chang}, H.-Y. and {Yi}, I. 2000, \apjl, 542, L17

\bibitem[{{Cool} {et~al.}(2008){Cool}, {Eisenstein}, {Hogg}, {Blanton},
  {Schlegel}, {Brinkmann}, {Lamb}, {Schneider}, \& {Berk}}]{ceh+08}
{Cool}, R.~J. {\it et al.} 2008, {GCN Circular} 7465

\bibitem[{{Covino} {et~al.}(2008){Covino}, {D'Avanzo}, {Fugazza}, {Antonelli},
  {Calzoletti}, {Campana}, {Chincarini}, {Conciatore}, {Cutini}, {D'Elia},
  {D'Alessio}, {Fiore}, {Goldoni}, {Guetta}, {Guidorzi}, {Israel}, {Maiorano},
  {Masetti}, {Melandri}, {Meurs}, {Nicastro}, {Palazzi}, {Pian}, {Piranomonte},
  {Stella}, {Stratta}, {Tagliaferri}, {Tosti}, {Testa}, {Vergani}, \&
  {Vitali}}]{Covino08_GCN7446}
{Covino}, S. {\it et al.} 2008, {GCN Circular} 7446

\bibitem[{{Cwiok} {et~al.}(2008{\natexlab{a}}){Cwiok}, {Dominik}, {Kasprowicz},
  {Majcher}, {Majczyna}, {Malek}, {Mankiewicz}, {Molak}, {Nawrocki},
  {Piotrowski}, {Rybka}, {Sokolowski}, {Uzycki}, {Wrochna}, \&
  {Zarnecki}}]{Cwiok08_GCN7445}
{Cwiok}, M. {\it et al.} 2008{\natexlab{a}}, {GCN Circular} 7445

\bibitem[{{Cwiok} {et~al.}(2008{\natexlab{b}}){Cwiok}, {Dominik}, {Kasprowicz},
  {Majcher}, {Majczyna}, {Malek}, {Mankiewicz}, {Molak}, {Nawrocki},
  {Piotrowski}, {Rybka}, {Sokolowski}, {Uzycki}, {Wrochna}, \&
  {Zarnecki}}]{Foley08}
---. 2008{\natexlab{b}}, {GCN Circular} 7445

\bibitem[{{Daigne} {et~al.}(2006){Daigne}, {Rossi}, \&
  {Mochkovitch}}]{2006MNRAS.372.1034D}
{Daigne}, F., {Rossi}, E.~M., and {Mochkovitch}, R. 2006, \mnras, 372, 1034

\bibitem[{{Eisner} {et~al.}(2007){Eisner}, {Hillenbrand}, {White}, {Bloom},
  {Akeson}, \& {Blake}}]{2007ApJ...669.1072E}
{Eisner}, J.~A., {Hillenbrand}, L.~A., {White}, R.~J., {Bloom}, J.~S.,
  {Akeson}, R.~L., and {Blake}, C.~H. 2007, \apj, 669, 1072

\bibitem[{{Ferrero} {et~al.}(2006){Ferrero}, {Kann}, {Zeh}, {Klose}, {Pian},
  {Palazzi}, {Masetti}, {Hartmann}, {Sollerman}, {Deng}, {Filippenko},
  {Greiner}, {Hughes}, {Mazzali}, {Li}, {Rol}, {Smith}, \&
  {Tanvir}}]{Ferrero+2006}
{Ferrero}, P. {\it et al.} 2006, \aap, 457, 857

\bibitem[{{Filippenko} {et~al.}(2001){Filippenko}, {Li}, {Treffers}, \&
  {Modjaz}}]{Filippenko2001}
{Filippenko}, A.~V., {Li}, W.~D., {Treffers}, R.~R., and {Modjaz}, M. 2001, in
  Astronomical Society of the Pacific Conference Series, Vol. 246, IAU Colloq.
  183: Small Telescope Astronomy on Global Scales, ed. B.~{Paczynski}, W.-P.
  {Chen}, \& C.~{Lemme} ({San Francisco, CA}: {ASP}), 121

\bibitem[{{Foley} {et~al.}(2006){Foley}, {Perley}, {Pooley}, {Prochaska},
  {Bloom}, {Li}, {Cobb}, {Chen}, {Aldering}, {Bailyn}, {Blake}, {Falco},
  {Green}, {Kowalski}, {Perlmutter}, {Roth}, \& {Volk}}]{Foley06}
{Foley}, R.~J. {\it et al.} 2006, \apj, 645, 450

\bibitem[{{Galama} {et~al.}(1999){Galama}, {Briggs}, {Wijers}, {Vreeswijk},
  {Rol}, {Band}, {Paradijs}, {Kouveliotou}, {Preece}, {Bremer}, {Smith},
  {Tilanus}, {Bruyn}, {Strom}, {Pooley}, {Castro-Tirado}, {Tanvir}, {Robinson},
  {Hurley}, {Heise}, {Telting}, {Rutten}, {Packham}, {Swaters}, {Davies},
  {Fassia}, {Green}, {Foster}, {Sagar}, {Pandey}, {Nilakshi}, {Yadav}, {Ofek},
  {Leibowitz}, {Ibbetson}, {Rhoads}, {Falco}, {Petry}, {Impey}, {Geballe}, \&
  {Bhattacharya}}]{gbw+99}
{Galama}, T.~J. {\it et al.} 1999, \nat, 398, 394

\bibitem[{{Gehrels} {et~al.}(2004){Gehrels}, {Chincarini}, {Giommi}, {Mason},
  {Nousek}, {Wells}, {White}, {Barthelmy}, {Burrows}, {Cominsky}, {Hurley},
  {Marshall}, {M{\'e}sz{\'a}ros}, {Roming}, {Angelini}, {Barbier}, {Belloni},
  {Campana}, {Caraveo}, {Chester}, {Citterio}, {Cline}, {Cropper}, {Cummings},
  {Dean}, {Feigelson}, {Fenimore}, {Frail}, {Fruchter}, {Garmire}, {Gendreau},
  {Ghisellini}, {Greiner}, {Hill}, {Hunsberger}, {Krimm}, {Kulkarni}, {Kumar},
  {Lebrun}, {Lloyd-Ronning}, {Markwardt}, {Mattson}, {Mushotzky}, {Norris},
  {Osborne}, {Paczynski}, {Palmer}, {Park}, {Parsons}, {Paul}, {Rees},
  {Reynolds}, {Rhoads}, {Sasseen}, {Schaefer}, {Short}, {Smale}, {Smith},
  {Stella}, {Tagliaferri}, {Takahashi}, {Tashiro}, {Townsley}, {Tueller},
  {Turner}, {Vietri}, {Voges}, {Ward}, {Willingale}, {Zerbi}, \&
  {Zhang}}]{Gehrels:2004p672}
{Gehrels}, N. {\it et al.} 2004, \apj, 611, 1005

\bibitem[{{Giblin} {et~al.}(1999){Giblin}, {van Paradijs}, {Kouveliotou},
  {Connaughton}, {Wijers}, {Briggs}, {Preece}, \& {Fishman}}]{Giblin99}
{Giblin}, T.~W., {van Paradijs}, J., {Kouveliotou}, C., {Connaughton}, V.,
  {Wijers}, R.~A.~M.~J., {Briggs}, M.~S., {Preece}, R.~D., and {Fishman}, G.~J.
  1999, \apjl, 524, L47

\bibitem[{{Golenetskii} {et~al.}(2008){Golenetskii}, {Aptekar}, {Mazets},
  {Pal'shin}, {Frederiks}, \& {Cline}}]{gam+08}
{Golenetskii}, S., {Aptekar}, R., {Mazets}, E., {Pal'shin}, V., {Frederiks},
  D., and {Cline}, T. 2008, {GCN Circular} 7482

\bibitem[{{Granot} \& {Kumar}(2006)}]{Granot06}
{Granot}, J. and {Kumar}, P. 2006, \mnras, 366, L13

\bibitem[{{Grindlay}(2007)}]{2007AIPC..921..211G}
{Grindlay}, J.~E. 2007, in American Institute of Physics Conference Series,
  Vol. 921, The First GLAST Symposium, ed. S.~{Ritz}, P.~{Michelson}, \& C.~A.
  {Meegan} ({New York, NY}: {AIP}), 211--216

\bibitem[{{Henden}(2008)}]{Henden08_GCN7528}
{Henden}, A. 2008, {GCN Circular} 7528

\bibitem[{{Hentunen} {et~al.}(2008){Hentunen}, {Oksanen}, \&
  {Kehusmaa}}]{Hentunen08_GCN7484}
{Hentunen}, V., {Oksanen}, A., and {Kehusmaa}, P. 2008, {GCN Circular} 7484

\bibitem[{{Hook} {et~al.}(2004){Hook}, {J{\o}rgensen}, {Allington-Smith},
  {Davies}, {Metcalfe}, {Murowinski}, \& {Crampton}}]{Hook04}
{Hook}, I.~M., {J{\o}rgensen}, I., {Allington-Smith}, J.~R., {Davies}, R.~L.,
  {Metcalfe}, N., {Murowinski}, R.~G., and {Crampton}, D. 2004, \pasp, 116, 425

\bibitem[{{Horne}(1986)}]{Horne86}
{Horne}, K. 1986, \pasp, 98, 609

\bibitem[{{Jakobsson} {et~al.}(2004){Jakobsson}, {Hjorth}, {Ramirez-Ruiz},
  {Kouveliotou}, {Pedersen}, {Fynbo}, {Gorosabel}, {Watson}, {Jensen}, {Grav},
  {Hansen}, {Michelsen}, {Andersen}, {Weidinger}, \&
  {Pedersen}}]{Jakobsson:2004p4601}
{Jakobsson}, P. {\it et al.} 2004, New Astronomy, 9, 435

\bibitem[{{Jakobsson} {et~al.}(2006){Jakobsson}, {Levan}, {Fynbo}, {Priddey},
  {Hjorth}, {Tanvir}, {Watson}, {Jensen}, {Sollerman}, {Natarajan},
  {Gorosabel}, {Castro Cer{\'o}n}, {Pedersen}, {Pursimo}, {{\'A}rnad{\'o}ttir},
  {Castro-Tirado}, {Davis}, {Deeg}, {Fiuza}, {Mykolaitis}, \&
  {Sousa}}]{2006A&A...447..897J}
---. 2006, \aap, 447, 897

\bibitem[{{Jelinek} {et~al.}(2008){Jelinek}, {Castro-Tirado}, {Chantry}, \&
  {Pl{\'a}}}]{Jelinek08_GCN7476}
{Jelinek}, M., {Castro-Tirado}, A.~J., {Chantry}, V., and {Pl{\'a}}, J. 2008,
  {GCN Circular} 7476

\bibitem[{{Kaneko} {et~al.}(2006){Kaneko}, {Preece}, {Briggs}, {Paciesas},
  {Meegan}, \& {Band}}]{Kaneko06}
{Kaneko}, Y., {Preece}, R.~D., {Briggs}, M.~S., {Paciesas}, W.~S., {Meegan},
  C.~A., and {Band}, D.~L. 2006, \apjs, 166, 298

\bibitem[{{Kann} {et~al.}(2006){Kann}, {Klose}, \& {Zeh}}]{Kann2006}
{Kann}, D.~A., {Klose}, S., and {Zeh}, A. 2006, \apj, 641, 993

\bibitem[{{Kann} {et~al.}(2007{\natexlab{a}}){Kann}, {Klose}, {Zhang},
  {Malesani}, {Nakar}, {Wilson}, {Butler}, {Antonelli}, {Chincarini}, {Cobb},
  {Covino}, {D'Avanzo}, {D'Elia}, {Della Valle}, {Ferrero}, {Fugazza},
  {Gorosabel}, {Israel}, {Mannucci}, {Piranomonte}, {Schulze}, {Stella},
  {Tagliaferri}, \& {Wiersema}}]{Kann:2006p3171}
{Kann}, D.~A. {\it et al.} 2007{\natexlab{a}}, ArXiv e-prints, 0712.2186

\bibitem[{{Kann} {et~al.}(2007{\natexlab{b}}){Kann}, {Masetti}, \&
  {Klose}}]{Kann:2007p133}
{Kann}, D.~A., {Masetti}, N., and {Klose}, S. 2007{\natexlab{b}}, \aj, 133,
  1187

\bibitem[{{Karpov} {et~al.}(2008{\natexlab{a}}){Karpov}, {Beskin}, {Bondar},
  {Bartolini}, {Greco}, {Guarnieri}, {Nanni}, {Piccioni}, {Terra}, {Molinari},
  {Chincarini}, {Zerbi}, {Covino}, {Testa}, {Tosti}, {Vitali}, {Antonelli},
  {Conconi}, {Cutispoto}, {Malaspina}, {Nicastro}, {Palazzi}, {Meurs}, \&
  {Goldoni}}]{Karpov08_GCN7452}
{Karpov}, S. {\it et al.} 2008{\natexlab{a}}, {GCN Circular} 7452

\bibitem[{{Karpov} {et~al.}(2008{\natexlab{b}}){Karpov}, {Beskin}, {Bondar},
  {Bartolini}, {Greco}, {Guarnieri}, {Piccioni}, {Nanni}, {Terra}, {Molinari},
  {Chincarini}, {Zerbi}, {Covino}, {Testa}, {Tosti}, {Vitali}, {Antonelli},
  {Conconi}, {Cutispoto}, {Malaspina}, {Nicastro}, {Palazzi}, {Meurs}, \&
  {Goldoni}}]{Karpov08_GCN7558}
---. 2008{\natexlab{b}}, {GCN Circular} 7558

\bibitem[{{Kobayashi}(2000)}]{Kobayashi2000}
{Kobayashi}, S. 2000, \apj, 545, 807

\bibitem[{{Krugly} {et~al.}(2008){Krugly}, {Slyusarev}, \&
  {Pozanenko}}]{Krugly08_GCN7519}
{Krugly}, Y., {Slyusarev}, I., and {Pozanenko}, A. 2008, {GCN Circular} 7519

\bibitem[{{Kumar} {et~al.}(2007){Kumar}, {McMahon}, {Panaitescu}, {Willingale},
  {O'Brien}, {Burrows}, {Cummings}, {Gehrels}, {Holland}, {Pandey}, {vanden
  Berk}, \& {Zane}}]{Kumar07}
{Kumar}, P. {\it et al.} 2007, \mnras, 376, L57

\bibitem[{{Kumar} \& {Panaitescu}(2008)}]{Kumar+2008}
{Kumar}, P. and {Panaitescu}, A. 2008, ArXiv e-prints, 0805.0144

\bibitem[{{Laursen} \& {Stanek}(2003)}]{Laursen:2003p4600}
{Laursen}, L.~T. and {Stanek}, K.~Z. 2003, \apjl, 597, L107

\bibitem[{{Lazzati} {et~al.}(2001){Lazzati}, {Ghisellini}, {Haardt}, \&
  {Fern{\'a}ndez-Soto}}]{2001grba.conf..191L}
{Lazzati}, D., {Ghisellini}, G., {Haardt}, F., and {Fern{\'a}ndez-Soto}, A.
  2001, in Gamma-ray Bursts in the Afterglow Era, ed. E.~{Costa},
  F.~{Frontera}, \& J.~{Hjorth}, 191

\bibitem[{{Le} \& {Dermer}(2007)}]{2007ApJ...661..394L}
{Le}, T. and {Dermer}, C.~D. 2007, \apj, 661, 394

\bibitem[{{Levan} {et~al.}(2008){Levan}, {Tanvir}, {Fruchter}, \&
  {Graham}}]{Levan08_GCN7710}
{Levan}, A.~J., {Tanvir}, N.~R., {Fruchter}, A.~S., and {Graham}, J. 2008, {GCN
  Circular} 7710

\bibitem[{{Li} {et~al.}(2008){Li}, {Chornock}, {Perley}, \&
  {Filippenko}}]{Li08_GCN7438}
{Li}, W., {Chornock}, R., {Perley}, D.~A., and {Filippenko}, A.~V. 2008, {GCN
  Circular} 7438

\bibitem[{{Li} {et~al.}(2003){Li}, {Filippenko}, {Chornock}, \&
  {Jha}}]{Li:2003p703}
{Li}, W., {Filippenko}, A.~V., {Chornock}, R., and {Jha}, S. 2003, \pasp, 115,
  844

\bibitem[{{Li} {et~al.}(2006){Li}, {Jha}, {Filippenko}, {Bloom}, {Pooley},
  {Foley}, \& {Perley}}]{Li:2006p4490}
{Li}, W., {Jha}, S., {Filippenko}, A.~V., {Bloom}, J.~S., {Pooley}, D.,
  {Foley}, R.~J., and {Perley}, D.~A. 2006, \pasp, 118, 37

\bibitem[{{Li} \& {Chevalier}(2003)}]{LiChevalier2003}
{Li}, Z.-Y. and {Chevalier}, R.~A. 2003, in Lecture Notes in Physics, Berlin
  Springer Verlag, Vol. 598, Supernovae and Gamma-Ray Bursters, ed.
  K.~{Weiler}, 419--444

\bibitem[{{Lipkin} {et~al.}(2004){Lipkin}, {Ofek}, {Gal-Yam}, {Leibowitz},
  {Poznanski}, {Kaspi}, {Polishook}, {Kulkarni}, {Fox}, {Berger}, {Mirabal},
  {Halpern}, {Bureau}, {Fathi}, {Price}, {Peterson}, {Frebel}, {Schmidt},
  {Orosz}, {Fitzgerald}, {Bloom}, {van Dokkum}, {Bailyn}, {Buxton}, \&
  {Barsony}}]{Lipkin:2004p552}
{Lipkin}, Y.~M. {\it et al.} 2004, \apj, 606, 381

\bibitem[{{Metzger} {et~al.}(2008){Metzger}, {Thompson}, \&
  {Quataert}}]{Metzger:2007p4684}
{Metzger}, B.~D., {Thompson}, T.~A., and {Quataert}, E. 2008, \apj, 676, 1130

\bibitem[{{Mundell} {et~al.}(2007){Mundell}, {Melandri}, {Guidorzi},
  {Kobayashi}, {Steele}, {Malesani}, {Amati}, {D'Avanzo}, {Bersier}, {Gomboc},
  {Rol}, {Bode}, {Carter}, {Mottram}, {Monfardini}, {Smith}, {Malhotra},
  {Wang}, {Bannister}, {O'Brien}, \& {Tanvir}}]{Mundell07}
{Mundell}, C.~G. {\it et al.} 2007, \apj, 660, 489

\bibitem[{{Naoz} \& {Bromberg}(2007)}]{Naoz+2007}
{Naoz}, S. and {Bromberg}, O. 2007, \mnras, 380, 757

\bibitem[{{Novak}(2008)}]{Novak08_GCN7504}
{Novak}, R. 2008, {GCN Circular} 7504

\bibitem[{{Oke} \& {Gunn}(1982)}]{oke82}
{Oke}, J.~B. and {Gunn}, J.~E. 1982, \pasp, 94, 586

\bibitem[{{Panaitescu} \& {Vestrand}(2008)}]{Panaitescu:2008p3168}
{Panaitescu}, A. and {Vestrand}, W.~T. 2008, \mnras, 387, 497

\bibitem[{{Perley} {et~al.}(2008){Perley}, {Bloom}, {Butler}, {Pollack},
  {Holtzman}, {Blake}, {Kocevski}, {Vestrand}, {Li}, {Foley}, {Bellm}, {Chen},
  {Prochaska}, {Starr}, {Filippenko}, {Falco}, {Szentgyorgyi}, {Wren},
  {Wozniak}, {White}, \& {Pergande}}]{Perley+2008a}
{Perley}, D.~A. {\it et al.} 2008, \apj, 672, 449

\bibitem[{{Podsiadlowski} {et~al.}(2004){Podsiadlowski}, {Mazzali}, {Nomoto},
  {Lazzati}, \& {Cappellaro}}]{Posiadlowski2004}
{Podsiadlowski}, P., {Mazzali}, P.~A., {Nomoto}, K., {Lazzati}, D., and
  {Cappellaro}, E. 2004, \apjl, 607, L17

\bibitem[{{Poole} {et~al.}(2008){Poole}, {Breeveld}, {Page}, {Landsman},
  {Holland}, {Roming}, {Kuin}, {Brown}, {Gronwall}, {Hunsberger}, {Koch},
  {Mason}, {Schady}, {vanden Berk}, {Blustin}, {Boyd}, {Broos}, {Carter},
  {Chester}, {Cucchiara}, {Hancock}, {Huckle}, {Immler}, {Ivanushkina},
  {Kennedy}, {Marshall}, {Morgan}, {Pandey}, {de Pasquale}, {Smith}, \&
  {Still}}]{Poole2008}
{Poole}, T.~S. {\it et al.} 2008, \mnras, 383, 627

\bibitem[{{Price} {et~al.}(2003){Price}, {Kulkarni}, {Berger}, {Fox}, {Bloom},
  {Djorgovski}, {Frail}, {Galama}, {Harrison}, {McCarthy}, {Reichart}, {Sari},
  {Yost}, {Jerjen}, {Flint}, {Phillips}, {Warren}, {Axelrod}, {Chevalier},
  {Holtzman}, {Kimble}, {Schmidt}, {Wheeler}, {Frontera}, {Costa}, {Piro},
  {Hurley}, {Cline}, {Guidorzi}, {Montanari}, {Mazets}, {Golenetskii},
  {Mitrofanov}, {Anfimov}, {Kozyrev}, {Litvak}, {Sanin}, {Boynton}, {Fellows},
  {Harshman}, {Shinohara}, {Gal-Yam}, {Ofek}, \& {Lipkin}}]{Price:2003p575}
{Price}, P.~A. {\it et al.} 2003, \apj, 589, 838

\bibitem[{{Racusin} {et~al.}(2008{\natexlab{a}}){Racusin}, {Gehrels},
  {Holland}, {Kennea}, {Markwardt}, {Pagani}, {Palmer}, \&
  {Stamatikos}}]{rgh+08}
{Racusin}, J.~L., {Gehrels}, N., {Holland}, S.~T., {Kennea}, J.~A.,
  {Markwardt}, C.~B., {Pagani}, C., {Palmer}, D.~M., and {Stamatikos}, M.
  2008{\natexlab{a}}, {GCN Circular} 7427

\bibitem[{{Racusin} {et~al.}(2008{\natexlab{b}}){Racusin}, {Karpov},
  {Sokolowski}, {Granot}, {Wu}, {Pal'shin}, {Covino}, {van der Horst}, {Oates},
  {Schady}, {Smith}, {Cummings}, {Starling}, {Piotrowski}, {Zhang}, {Evans},
  {Holland}, {Malek}, {Page}, {Vetere}, {Margutti}, {Guidorzi}, {Kamble},
  {Curran}, {Beardmore}, {Kouveliotou}, {Mankiewicz}, {Melandri}, {O'Brien},
  {Page}, {Piran}, {Tanvir}, {Wrochna}, {Aptekar}, {Bartolini}, {Barthelmy},
  {Beskin}, {Bondar}, {Campana}, {Cucchiara}, {Cwiok}, {D'Avanzo}, {D'Elia},
  {Della Valle}, {Dominik}, {Falcone}, {Fiore}, {Fox}, {Frederiks}, {Fruchter},
  {Fugazza}, {Garrett}, {Gehrels}, {Golenetskii}, {Gomboc}, {Greco},
  {Guarnieri}, {Immler}, {Kasprowicz}, {Levan}, {Mazets}, {Molinari},
  {Moretti}, {Nawrocki}, {Oleynik}, {Osborne}, {Pagani}, {Paragi}, {Perri},
  {Piccioni}, {Ramirez-Ruiz}, {Roming}, {Steele}, {Strom}, {Testa}, {Tosti},
  {Ulanov}, {Wiersema}, {Wijers}, {Zarnecki}, {Zerbi}, {Meszaros},
  {Chincarini}, \& {Burrows}}]{Racusineteveryone}
{Racusin}, J.~L. {\it et al.} 2008{\natexlab{b}}, ArXiv e-prints, 0805.1557

\bibitem[{{Sari} {et~al.}(1999){Sari}, {Piran}, \& {Halpern}}]{Sari1999}
{Sari}, R., {Piran}, T., and {Halpern}, J.~P. 1999, \apjl, 519, L17

\bibitem[{{Sari} {et~al.}(1998){Sari}, {Piran}, \& {Narayan}}]{Sari+1998}
{Sari}, R., {Piran}, T., and {Narayan}, R. 1998, \apjl, 497, L17+

\bibitem[{{Schady} {et~al.}(2007{\natexlab{a}}){Schady}, {de Pasquale}, {Page},
  {Vetere}, {Pandey}, {Wang}, {Cummings}, {Zhang}, {Zane}, {Breeveld},
  {Burrows}, {Gehrels}, {Gronwall}, {Hunsberger}, {Markwardt}, {Mason},
  {M{\'e}sz{\'a}ros}, {Norris}, {Oates}, {Pagani}, {Poole}, {Roming}, {Smith},
  \& {vanden Berk}}]{Schady07}
{Schady}, P. {\it et al.} 2007{\natexlab{a}}, \mnras, 380, 1041

\bibitem[{{Schady} {et~al.}(2007{\natexlab{b}}){Schady}, {Mason}, {Page}, {de
  Pasquale}, {Morris}, {Romano}, {Roming}, {Immler}, \& {vanden
  Berk}}]{Schady+2007}
---. 2007{\natexlab{b}}, \mnras, 377, 273

\bibitem[{{Schlegel} {et~al.}(1998){Schlegel}, {Finkbeiner}, \&
  {Davis}}]{Schlegel:1998p695}
{Schlegel}, D.~J., {Finkbeiner}, D.~P., and {Davis}, M. 1998, \apj, 500, 525

\bibitem[{{Schneider} {et~al.}(2007){Schneider}, {Hall}, {Richards}, {Strauss},
  {Vanden Berk}, {Anderson}, {Brandt}, {Fan}, {Jester}, {Gray}, {Gunn},
  {SubbaRao}, {Thakar}, {Stoughton}, {Szalay}, {Yanny}, {York}, {Bahcall},
  {Barentine}, {Blanton}, {Brewington}, {Brinkmann}, {Brunner}, {Castander},
  {Csabai}, {Frieman}, {Fukugita}, {Harvanek}, {Hogg}, {Ivezi{\'c}}, {Kent},
  {Kleinman}, {Knapp}, {Kron}, {Krzesi{\'n}ski}, {Long}, {Lupton}, {Nitta},
  {Pier}, {Saxe}, {Shen}, {Snedden}, {Weinberg}, \& {Wu}}]{Schneider:2007p4602}
{Schneider}, D.~P. {\it et al.} 2007, \aj, 134, 102

\bibitem[{{Schubel} {et~al.}(2008){Schubel}, {Reichart}, {Nysewander},
  {LaCluyze}, {Ivarsen}, {Crain}, {Foster}, {Brennan}, {Haislip}, {Styblova},
  \& {Trotter}}]{Schubel08_GCN7461}
{Schubel}, M. {\it et al.} 2008, {GCN Circular} 7461

\bibitem[{{Smith} {et~al.}(2007){Smith}, {Li}, {Foley}, {Wheeler}, {Pooley},
  {Chornock}, {Filippenko}, {Silverman}, {Quimby}, {Bloom}, \&
  {Hansen}}]{Smith:2007p532}
{Smith}, N. {\it et al.} 2007, \apj, 666, 1116

\bibitem[{{Stanek} {et~al.}(2005){Stanek}, {Garnavich}, {Nutzman}, {Hartman},
  {Garg}, {Adelberger}, {Berlind}, {Bonanos}, {Calkins}, {Challis}, {Gaudi},
  {Holman}, {Kirshner}, {McLeod}, {Osip}, {Pimenova}, {Reiprich}, {Romanishin},
  {Spahr}, {Tegler}, \& {Zhao}}]{Stanek2005}
{Stanek}, K.~Z. {\it et al.} 2005, \apjl, 626, L5

\bibitem[{{Stanek} {et~al.}(2006){Stanek}, {Gnedin}, {Beacom}, {Gould},
  {Johnson}, {Kollmeier}, {Modjaz}, {Pinsonneault}, {Pogge}, \&
  {Weinberg}}]{Stanek2006}
---. 2006, Acta Astronomica, 56, 333

\bibitem[{{Swan} {et~al.}(2008){Swan}, {Yuan}, \&
  {Rujopakarn}}]{Swan08_GCN7470}
{Swan}, H., {Yuan}, F., and {Rujopakarn}, W. 2008, {GCN Circular} 7470

\bibitem[{{Tagliaferri} {et~al.}(2005){Tagliaferri}, {Antonelli}, {Chincarini},
  {Fern{\'a}ndez-Soto}, {Malesani}, {Della Valle}, {D'Avanzo}, {Grazian},
  {Testa}, {Campana}, {Covino}, {Fiore}, {Stella}, {Castro-Tirado},
  {Gorosabel}, {Burrows}, {Capalbi}, {Cusumano}, {Conciatore}, {D'Elia},
  {Filliatre}, {Fugazza}, {Gehrels}, {Goldoni}, {Guetta}, {Guziy}, {Held},
  {Hurley}, {Israel}, {Jel{\'{\i}}nek}, {Lazzati}, {L{\'o}pez-Echarri},
  {Melandri}, {Mirabel}, {Moles}, {Moretti}, {Mason}, {Nousek}, {Osborne},
  {Pellizza}, {Perna}, {Piranomonte}, {Piro}, {de Ugarte Postigo}, \&
  {Romano}}]{Tagliaferri+2005}
{Tagliaferri}, G. {\it et al.} 2005, \aap, 443, L1

\bibitem[{{Tanvir} {et~al.}(2008){Tanvir}, {Levan}, {Fruchter}, {Graham},
  {Wiersema}, \& {Rol}}]{Tanvir08_GCN7569}
{Tanvir}, N.~R., {Levan}, A.~J., {Fruchter}, A.~S., {Graham}, J., {Wiersema},
  K., and {Rol}, E. 2008, {GCN Circular} 7569

\bibitem[{{Updike} {et~al.}(2008){Updike}, {Haislip}, {Nysewander}, {Fruchter},
  {Kann}, {Klose}, {Milne}, {Williams}, {Zheng}, {Hergenrother}, {Prochaska},
  {Halpern}, {Mirabal}, {Thorstensen}, {van der Horst}, {Starling}, {Racusin},
  {Burrows}, {Kuin}, {Roming}, {Bellm}, {Hurley}, {Li}, {Filippenko}, {Blake},
  {Starr}, {Falco}, {Brown}, {Dai}, {Deng}, {Xin}, {Qiu}, {Wei}, {Urata},
  {Nanni}, {Maiorano}, {Palazzi}, {Greco}, {Bartolini}, {Guarnieri},
  {Piccioni}, {Pizzichini}, {Terra}, {Misra}, {Bhatt}, {Anupama}, {Fan},
  {Jiang}, {Wijers}, {Reichart}, {Eid}, {Bryngelson}, {Puls}, {Goldthwaite}, \&
  {Hartmann}}]{Updike+2008}
{Updike}, A.~C. {\it et al.} 2008, ArXiv e-prints, 0805.1094

\bibitem[{{Vreeswijk} {et~al.}(2008){Vreeswijk}, {Smette}, {Malesani}, {Fynbo},
  {Milvang-Jensen}, {Jakobsson}, {Jaunsen}, \& {Ledoux}}]{Vreeswijk08}
{Vreeswijk}, P.~M., {Smette}, A., {Malesani}, D., {Fynbo}, J.~P.~U.,
  {Milvang-Jensen}, B., {Jakobsson}, P., {Jaunsen}, A.~O., and {Ledoux}, C.
  2008, {GCN Circular} 7444

\bibitem[{{Wood-Vasey} {et~al.}(2007){Wood-Vasey}, {Friedman}, {Bloom},
  {Hicken}, {Modjaz}, {Kirshner}, {Starr}, {Blake}, {Falco}, {Szentgyorgyi},
  {Challis}, {Blondin}, \& {Rest}}]{wfb+08}
{Wood-Vasey}, W.~M. {\it et al.} 2007, ArXiv e-prints, 0711.2068

\bibitem[{{Wozniak} {et~al.}(2008){Wozniak}, {Vestrand}, {Wren}, \&
  {Davis}}]{Wozniak08_GCN7464}
{Wozniak}, P., {Vestrand}, W.~T., {Wren}, J., and {Davis}, H. 2008, {GCN
  Circular} 7464

\bibitem[{{Zeh} {et~al.}(2004){Zeh}, {Klose}, \& {Hartmann}}]{Zeh+2004}
{Zeh}, A., {Klose}, S., and {Hartmann}, D.~H. 2004, \apj, 609, 952

\end{thebibliography}


\begin{deluxetable}{rccll}
\tabletypesize{\small}
\tablecaption{PAIRITEL Observations of GRB\,080319B\label{tab:photlogptel}}
\tablecolumns{6}
\tablehead{
\colhead{$t_{\rm mid}$\tablenotemark{a}} & \colhead{Filter} &
\colhead{Exp.~time} &
\colhead{Mag.\tablenotemark{b}} & \colhead{Flux\tablenotemark{b}} \\
\colhead{sec} & \colhead{} &
\colhead{sec} & \colhead{} &
\colhead{$\mu$Jy}}
\startdata
  344.0 & {\rm J}     &    7.8 & $10.389 \pm 0.048$ & $111255.0\pm  4815.5$ \\
  352.3 & {\rm J}     &    7.8 & $10.483 \pm 0.048$ & $102084.5\pm  4419.0$ \\
  360.2 & {\rm J}     &    7.8 & $10.563 \pm 0.048$ & $ 94763.2\pm  4102.4$ \\
  380.1 & {\rm J}     &    7.8 & $10.652 \pm 0.048$ & $ 87313.1\pm  3779.8$ \\
  388.4 & {\rm J}     &    7.8 & $10.722 \pm 0.048$ & $ 81846.4\pm  3544.3$ \\
  396.3 & {\rm J}     &    7.8 & $10.781 \pm 0.048$ & $ 77538.9\pm  3357.8$ \\
  416.3 & {\rm J}     &    7.8 & $10.877 \pm 0.048$ & $ 70983.9\pm  3074.6$ \\
  424.5 & {\rm J}     &    7.8 & $10.896 \pm 0.048$ & $ 69752.5\pm  3021.6$ \\
  432.4 & {\rm J}     &    7.8 & $10.952 \pm 0.048$ & $ 66227.7\pm  2868.9$ \\
  344.0 & {\rm H}     &    7.8 & $ 9.639 \pm 0.042$ & $142696.2\pm  5421.2$ \\
  352.3 & {\rm H}     &    7.8 & $ 9.721 \pm 0.042$ & $132364.7\pm  5029.4$ \\
  360.2 & {\rm H}     &    7.8 & $ 9.753 \pm 0.042$ & $128513.3\pm  4883.6$ \\
  380.1 & {\rm H}     &    7.8 & $ 9.946 \pm 0.042$ & $107582.1\pm  4088.8$ \\
  388.4 & {\rm H}     &    7.8 & $ 9.985 \pm 0.042$ & $103750.1\pm  3944.2$ \\
  396.3 & {\rm H}     &    7.8 & $10.115 \pm 0.042$ & $ 92078.9\pm  3501.5$ \\
  416.3 & {\rm H}     &    7.8 & $10.127 \pm 0.042$ & $ 91066.8\pm  3464.2$ \\
  424.5 & {\rm H}     &    7.8 & $10.177 \pm 0.042$ & $ 86920.1\pm  3306.4$ \\
  432.4 & {\rm H}     &    7.8 & $10.245 \pm 0.042$ & $ 81665.8\pm  3107.6$ \\
  344.0 & {$\rm K_s$} &    7.8 & $ 9.067 \pm 0.058$\tablenotemark{x} & $157495.6\pm  8197.4$\tablenotemark{x} \\
  352.3 & {$\rm K_s$} &    7.8 & $ 9.308 \pm 0.058$\tablenotemark{x} & $126102.6\pm  6564.2$\tablenotemark{x} \\
  360.2 & {$\rm K_s$} &    7.8 & $ 9.207 \pm 0.058$\tablenotemark{x} & $138421.8\pm  7205.9$\tablenotemark{x} \\
  380.1 & {$\rm K_s$} &    7.8 & $ 9.312 \pm 0.058$ & $125650.6\pm  6541.1$ \\
  388.4 & {$\rm K_s$} &    7.8 & $ 9.310 \pm 0.058$ & $125949.5\pm  6557.6$ \\
  396.3 & {$\rm K_s$} &    7.8 & $ 9.419 \pm 0.058$ & $113840.4\pm  5927.9$ \\
  416.3 & {$\rm K_s$} &    7.8 & $ 9.508 \pm 0.058$ & $104946.6\pm  5465.3$ \\
  424.5 & {$\rm K_s$} &    7.8 & $ 9.546 \pm 0.058$ & $101331.5\pm  5277.5$ \\
  432.4 & {$\rm K_s$} &    7.8 & $ 9.638 \pm 0.058$ & $ 93031.2\pm  4845.6$ \\
\enddata 
\tablecomments{Because of the very large number of exposures acquired, only a few representative points are given.  The full table of photometry containing all 406 points is available online.  Includes only non-saturated exposures; photometry from the saturated epochs is given in Tables \ref{tab:shortr} and \ref{tab:annulus}.}
\tablenotetext{a}{Exposure mid-time, measured from the \emph{Swift} trigger (UTC 06:12:49).}
\tablenotetext{b}{Observed value; not corrected for Galactic extinction.}
\tablenotetext{x}{Point not used in modeling.}
\end{deluxetable}

\begin{deluxetable}{lccll}
\tabletypesize{\small}
\tablecaption{PAIRITEL Short-Read Observations of GRB\,080319B During Saturated Regime \label{tab:shortr}}
\tablecolumns{6}
\tablehead{
\colhead{$t_{\rm mid}$\tablenotemark{a}} & \colhead{Filter} &
\colhead{Exp.~time} &
\colhead{Mag.\tablenotemark{b}} & \colhead{Flux\tablenotemark{b}} \\
\colhead{sec} & \colhead{} &
\colhead{sec} & \colhead{} &
\colhead{$\mu$Jy}}
\startdata
   59.3 & {\rm J}     &  0.051 & $ 5.206 \pm 0.055$ & $1.32\times 10^{7}\pm6.52\times 10^{5}$ \\
   67.3 & {\rm J}     &  0.051 & $ 5.627 \pm 0.055$ & $8.94\times 10^{6}\pm4.43\times 10^{5}$ \\
   87.3 & {\rm J}     &  0.051 & $ 6.470 \pm 0.055$ & $4.11\times 10^{6}\pm2.04\times 10^{5}$ \\
   95.5 & {\rm J}     &  0.051 & $ 6.847 \pm 0.056$ & $2.91\times 10^{6}\pm1.45\times 10^{5}$ \\
  103.5 & {\rm J}     &  0.051 & $ 6.954 \pm 0.056$ & $2.63\times 10^{6}\pm1.32\times 10^{5}$ \\
   59.3 & {\rm H}     &  0.051 & $ 4.520 \pm 0.052$ & $1.59\times 10^{7}\pm7.47\times 10^{5}$ \\
   67.3 & {\rm H}     &  0.051 & $ 4.853 \pm 0.052$ & $1.17\times 10^{7}\pm5.50\times 10^{5}$ \\
   87.3 & {\rm H}     &  0.051 & $ 5.713 \pm 0.052$ & $5.30\times 10^{6}\pm2.50\times 10^{5}$ \\
   95.5 & {\rm H}     &  0.051 & $ 6.101 \pm 0.052$ & $3.71\times 10^{6}\pm1.75\times 10^{5}$ \\
  103.5 & {\rm H}     &  0.051 & $ 6.225 \pm 0.052$ & $3.31\times 10^{6}\pm1.56\times 10^{5}$ \\
   67.3 & {$\rm K_s$} &  0.051 & $ 4.362 \pm 0.048$ & $1.20\times 10^{7}\pm5.20\times 10^{5}$ \\
   87.3 & {$\rm K_s$} &  0.051 & $ 5.192 \pm 0.048$ & $5.59\times 10^{6}\pm2.43\times 10^{5}$ \\
   95.5 & {$\rm K_s$} &  0.051 & $ 5.308 \pm 0.048$ & $5.02\times 10^{6}\pm2.18\times 10^{5}$ \\
  103.5 & {$\rm K_s$} &  0.051 & $ 5.675 \pm 0.048$ & $3.58\times 10^{6}\pm1.56\times 10^{5}$ \\
\enddata 
\tablenotetext{a}{Exposure mid-time, measured from the \emph{Swift} trigger (UT 06:12:49).}
\tablenotetext{b}{Observed value; not corrected for Galactic extinction.}
\tablenotetext{x}{No saturated values were used in modeling.}
\tablecomments{Because of the large number of exposures acquired, only a few representative points are given.  The full table of photometry containing all 406 points is available online.}
\end{deluxetable}

\begin{deluxetable}{lccll}
\tabletypesize{\small}
\tablecaption{PAIRITEL Annulus Photometry of GRB\,080319B During Saturated Regime\label{tab:annulus}}
\tablecolumns{6}
\tablehead{
\colhead{$t_{\rm mid}$\tablenotemark{a}} & \colhead{Filter} &
\colhead{Exp.~time} &
\colhead{Mag.\tablenotemark{b}} & \colhead{Flux\tablenotemark{b}} \\
\colhead{sec} & \colhead{} &
\colhead{sec} & \colhead{} &
\colhead{$\mu$Jy}}
\startdata
   55.0 & {\rm J}     &    7.8 & $ 4.595 \pm 0.133$ & $2.31\times 10^{7}\pm2.66\times 10^{6}$ \\
   63.2 & {\rm J}     &    7.8 & $ 5.070 \pm 0.133$ & $1.49\times 10^{7}\pm1.72\times 10^{6}$ \\
   71.2 & {\rm J}     &    7.8 & $ 5.560 \pm 0.133$ & $9.51\times 10^{6}\pm1.10\times 10^{6}$ \\
   91.2 & {\rm J}     &    7.8 & $ 6.572 \pm 0.133$ & $3.74\times 10^{6}\pm4.31\times 10^{5}$ \\
   99.4 & {\rm J}     &    7.8 & $ 6.710 \pm 0.133$ & $3.30\times 10^{6}\pm3.80\times 10^{5}$ \\
  107.4 & {\rm J}     &    7.8 & $ 7.004 \pm 0.133$ & $2.51\times 10^{6}\pm2.90\times 10^{5}$ \\
   55.0 & {\rm H}     &    7.8 & $ 4.387 \pm 0.191$ & $1.80\times 10^{7}\pm2.90\times 10^{6}$ \\
   63.2 & {\rm H}     &    7.8 & $ 4.805 \pm 0.191$ & $1.22\times 10^{7}\pm1.97\times 10^{6}$ \\
   71.2 & {\rm H}     &    7.8 & $ 5.159 \pm 0.191$ & $8.84\times 10^{6}\pm1.42\times 10^{6}$ \\
   91.2 & {\rm H}     &    7.8 & $ 5.818 \pm 0.191$ & $4.82\times 10^{6}\pm7.76\times 10^{5}$ \\
   99.4 & {\rm H}     &    7.8 & $ 6.386 \pm 0.191$ & $2.86\times 10^{6}\pm4.60\times 10^{5}$ \\
  107.4 & {\rm H}     &    7.8 & $ 6.421 \pm 0.191$ & $2.76\times 10^{6}\pm4.45\times 10^{5}$ \\
   55.0 & {$\rm K_s$} &    7.8 & $ 3.853 \pm 0.463$ & $1.92\times 10^{7}\pm6.65\times 10^{6}$ \\
   63.2 & {$\rm K_s$} &    7.8 & $ 4.200 \pm 0.463$ & $1.39\times 10^{7}\pm4.83\times 10^{6}$ \\
   71.2 & {$\rm K_s$} &    7.8 & $ 4.510 \pm 0.463$ & $1.05\times 10^{7}\pm3.63\times 10^{6}$ \\
   91.2 & {$\rm K_s$} &    7.8 & $ 5.130 \pm 0.463$ & $5.91\times 10^{6}\pm2.05\times 10^{6}$ \\
   99.4 & {$\rm K_s$} &    7.8 & $ 5.659 \pm 0.463$ & $3.63\times 10^{6}\pm1.26\times 10^{6}$ \\
  107.4 & {$\rm K_s$} &    7.8 & $ 5.812 \pm 0.463$ & $3.16\times 10^{6}\pm1.09\times 10^{6}$ \\
\enddata 
\tablenotetext{a}{Exposure mid-time, measured from the \emph{Swift} trigger (UT 06:12:49).}
\tablenotetext{b}{Observed value; not corrected for Galactic extinction.}
\tablenotetext{x}{No saturated values were used in modeling.}
\tablecomments{Because of the large number of exposures acquired, only a few representative points are given.  The full table of photometry containing all 406 points is available online.}
\end{deluxetable}

\begin{deluxetable}{rccll}
\tabletypesize{\small}
\tablecaption{KAIT Observations of GRB\,080319B\label{tab:photlogkait}}
\tablecolumns{6}
\tablehead{
\colhead{$t_{\rm mid}$\tablenotemark{a}} & \colhead{Filter} &
\colhead{Exp.~time} &
\colhead{Mag.\tablenotemark{b}} & \colhead{Flux\tablenotemark{b}} \\
\colhead{sec} & \colhead{} &
\colhead{sec} & \colhead{} &
\colhead{$\mu$Jy}}
\startdata
 1217.0 & {\rm clear} &   20.0 & $14.116 \pm 0.008$ & $  7008.1\pm    51.4$ \\
 1308.0 & {\rm clear} &   20.0 & $14.261 \pm 0.007$ & $  6132.0\pm    39.4$ \\
 1404.0 & {\rm clear} &   20.0 & $14.424 \pm 0.011$ & $  5277.2\pm    53.2$ \\
 1495.0 & {\rm clear} &   20.0 & $14.548 \pm 0.013$ & $  4707.6\pm    56.0$ \\
 1586.0 & {\rm clear} &   20.0 & $14.651 \pm 0.011$ & $  4281.5\pm    43.2$ \\
 2098.0 & {\rm B}     &   20.0 & $15.623 \pm 0.045$ & $  2327.0\pm    94.5$ \\
 2249.0 & {\rm B}     &   20.0 & $15.750 \pm 0.041$ & $  2070.1\pm    76.7$ \\
 2373.0 & {\rm B}     &   20.0 & $15.814 \pm 0.042$ & $  1951.6\pm    74.1$ \\
 2546.5 & {\rm B}     &   40.0 & $15.911 \pm 0.037$ & $  1784.8\pm    59.8$ \\
 2791.0 & {\rm B}     &   40.0 & $16.094 \pm 0.035$ & $  1508.0\pm    47.8$ \\
 1157.0 & {\rm V}     &   20.0 & $14.345 \pm 0.022$ & $  6745.3\pm   135.3$ \\
 1248.0 & {\rm V}     &   20.0 & $14.442 \pm 0.022$ & $  6168.8\pm   123.7$ \\
 1339.0 & {\rm V}     &   20.0 & $14.651 \pm 0.022$ & $  5088.6\pm   102.1$ \\
 1435.0 & {\rm V}     &   20.0 & $14.815 \pm 0.025$ & $  4375.2\pm    99.6$ \\
 1526.0 & {\rm V}     &   20.0 & $14.939 \pm 0.026$ & $  3903.0\pm    92.4$ \\
 1188.0 & {\rm I}     &   20.0 & $13.777 \pm 0.013$ & $  7502.4\pm    89.3$ \\
 1279.0 & {\rm I}     &   20.0 & $13.879 \pm 0.012$ & $  6829.7\pm    75.1$ \\
 1370.0 & {\rm I}     &   20.0 & $14.016 \pm 0.015$ & $  6020.0\pm    82.6$ \\
 1465.0 & {\rm I}     &   20.0 & $14.176 \pm 0.020$ & $  5195.2\pm    94.8$ \\
 1557.0 & {\rm I}     &   20.0 & $14.328 \pm 0.018$ & $  4516.5\pm    74.3$ \\
\enddata 
\tablecomments{Because of the very large number of exposures acquired, only the first five exposures in each band are given.  The full table of photometry is available online.}
\tablenotetext{a}{Exposure mid-time, measured from the \emph{Swift} trigger (UTC 06:12:49).}
\tablenotetext{b}{Observed value; not corrected for Galactic extinction.}
\end{deluxetable}

\begin{deluxetable}{rccll}
\tabletypesize{\small}
\tablecaption{Nickel Observations of GRB\,080319B\label{tab:photlognickel}}
\tablecolumns{6}
\tablehead{
\colhead{$t_{\rm mid}$\tablenotemark{a}} & \colhead{Filter} &
\colhead{Exp.~time} &
\colhead{Mag.\tablenotemark{b}} & \colhead{Flux\tablenotemark{b}} \\
\colhead{sec} & \colhead{} &
\colhead{sec} & \colhead{} &
\colhead{$\mu$Jy}}
\startdata
 7154.7 & {\rm B}     &  360.0 & $17.365 \pm 0.005$ & $   467.7\pm     2.1$ \\
 8579.2 & {\rm B}     &  360.0 & $17.646 \pm 0.006$ & $   361.1\pm     2.0$ \\
10013.4 & {\rm B}     &  360.0 & $17.866 \pm 0.008$ & $   294.8\pm     2.2$ \\
11512.3 & {\rm B}     &  360.0 & $18.090 \pm 0.008$ & $   239.9\pm     1.8$ \\
13015.9 & {\rm B}     &  360.0 & $18.248 \pm 0.011$ & $   207.4\pm     2.1$ \\
14473.6 & {\rm B}     &  360.0 & $18.389 \pm 0.011$ & $   182.1\pm     1.8$ \\
16436.8 & {\rm B}     &  360.0 & $18.587 \pm 0.015$ & $   151.8\pm     2.1$ \\
17937.4 & {\rm B}     &  360.0 & $18.770 \pm 0.018$ & $   128.2\pm     2.1$ \\
19398.1 & {\rm B}     &  360.0 & $18.858 \pm 0.015$ & $   118.2\pm     1.6$ \\
20965.3 & {\rm B}     &  360.0 & $18.970 \pm 0.023$ & $   106.7\pm     2.2$ \\
 7444.0 & {\rm V}     &  300.0 & $17.251 \pm 0.005$ & $   464.1\pm     2.1$ \\
 8860.5 & {\rm V}     &  300.0 & $17.484 \pm 0.007$ & $   374.5\pm     2.4$ \\
10291.9 & {\rm V}     &  300.0 & $17.727 \pm 0.007$ & $   299.4\pm     1.9$ \\
11829.3 & {\rm V}     &  300.0 & $17.899 \pm 0.007$ & $   255.5\pm     1.6$ \\
13294.2 & {\rm V}     &  300.0 & $18.062 \pm 0.007$ & $   219.9\pm     1.4$ \\
14767.2 & {\rm V}     &  300.0 & $18.236 \pm 0.014$ & $   187.3\pm     2.4$ \\
16714.4 & {\rm V}     &  300.0 & $18.424 \pm 0.012$ & $   157.5\pm     1.7$ \\
18236.5 & {\rm V}     &  300.0 & $18.559 \pm 0.016$ & $   139.1\pm     2.0$ \\
19700.4 & {\rm V}     &  300.0 & $18.637 \pm 0.013$ & $   129.5\pm     1.5$ \\
22996.8 & {\rm V}     &  300.0 & $18.902 \pm 0.021$ & $   101.4\pm     1.9$ \\
 7782.0 & {\rm R}     &  300.0 & $16.991 \pm 0.006$ & $   496.1\pm     2.7$ \\
 9196.4 & {\rm R}     &  300.0 & $17.239 \pm 0.005$ & $   394.8\pm     1.8$ \\
10635.4 & {\rm R}     &  300.0 & $17.435 \pm 0.008$ & $   329.6\pm     2.4$ \\
12181.0 & {\rm R}     &  300.0 & $17.637 \pm 0.009$ & $   273.7\pm     2.3$ \\
13652.0 & {\rm R}     &  300.0 & $17.799 \pm 0.007$ & $   235.7\pm     1.5$ \\
15110.2 & {\rm R}     &  300.0 & $17.917 \pm 0.007$ & $   211.4\pm     1.4$ \\
17071.9 & {\rm R}     &  300.0 & $18.107 \pm 0.011$ & $   177.5\pm     1.8$ \\
18582.1 & {\rm R}     &  300.0 & $18.218 \pm 0.008$ & $   160.3\pm     1.2$ \\
20035.8 & {\rm R}     &  300.0 & $18.308 \pm 0.017$ & $   147.5\pm     2.3$ \\
23331.8 & {\rm R}     &  300.0 & $18.541 \pm 0.021$ & $   119.0\pm     2.3$ \\
 8122.0 & {\rm I}     &  300.0 & $16.797 \pm 0.004$ & $   464.7\pm     1.7$ \\
 9531.2 & {\rm I}     &  300.0 & $17.013 \pm 0.005$ & $   380.9\pm     1.7$ \\
10995.4 & {\rm I}     &  300.0 & $17.189 \pm 0.005$ & $   323.9\pm     1.5$ \\
12520.7 & {\rm I}     &  300.0 & $17.394 \pm 0.005$ & $   268.2\pm     1.2$ \\
14003.3 & {\rm I}     &  300.0 & $17.537 \pm 0.009$ & $   235.1\pm     1.9$ \\
15973.2 & {\rm I}     &  300.0 & $17.704 \pm 0.008$ & $   201.6\pm     1.5$ \\
17412.9 & {\rm I}     &  300.0 & $17.817 \pm 0.011$ & $   181.6\pm     1.8$ \\
18936.7 & {\rm I}     &  300.0 & $17.921 \pm 0.010$ & $   165.0\pm     1.5$ \\
20447.9 & {\rm I}     &  300.0 & $18.045 \pm 0.014$ & $   147.2\pm     1.9$ \\
23665.5 & {\rm I}     &  300.0 & $18.111 \pm 0.118$ & $   138.5\pm    14.3$ \\
\enddata 
\tablenotetext{a}{Exposure mid-time, measured from the \emph{Swift} trigger (UTC 06:12:49).}
\tablenotetext{b}{Observed value; not corrected for Galactic extinction.}
\end{deluxetable}

\begin{deluxetable}{rccll}
\tabletypesize{\small}
\tablecaption{UVOT Observations of GRB\,080319B\label{tab:photloguvot}}
\tablecolumns{6}
\tablehead{
\colhead{$t_{\rm mid}$\tablenotemark{a}} & \colhead{Filter} &
\colhead{Exp.~time} &
\colhead{Mag.\tablenotemark{b}} & \colhead{Flux\tablenotemark{b}} \\
\colhead{sec} & \colhead{} &
\colhead{sec} & \colhead{} &
\colhead{$\mu$Jy}}
\startdata
 6156.0 & {\rm UVW1}  &  196.6 & $16.146 \pm 0.043$ & $   343.2\pm    13.3$ \\
17398.0 & {\rm UVW1}  &  885.6 & $17.645 \pm 0.042$ & $    86.3\pm     3.3$ \\
28047.0 & {\rm UVW1}  &  427.9 & $18.316 \pm 0.086$ & $    46.5\pm     3.5$ \\
45393.0 & {\rm UVW1}  &  427.9 & $18.909 \pm 0.121$ & $    26.9\pm     2.8$ \\
75628.0 & {\rm UVW1}  &  200.6 & $19.266 \pm 0.213$ & $    19.4\pm     3.5$ \\
 5542.0 & {\rm UVW2}  &  196.6 & $16.192 \pm 0.046$ & $   239.6\pm     9.9$ \\
10709.0 & {\rm UVW2}  &  885.6 & $17.231 \pm 0.036$ & $    92.0\pm     3.0$ \\
22273.0 & {\rm UVW2}  &  266.3 & $18.394 \pm 0.116$ & $    31.5\pm     3.2$ \\
28931.0 & {\rm UVW2}  &  856.1 & $18.631 \pm 0.073$ & $    25.3\pm     1.6$ \\
46279.0 & {\rm UVW2}  &  856.1 & $19.228 \pm 0.102$ & $    14.6\pm     1.3$ \\
76050.0 & {\rm UVW2}  &  401.3 & $19.888 \pm 0.220$ & $     8.0\pm     1.5$ \\
 5951.0 & {\rm UVM2}  &  196.6 & $15.923 \pm 0.052$ & $   284.4\pm    13.3$ \\
12528.0 & {\rm UVM2}  &  334.6 & $17.144 \pm 0.071$ & $    92.4\pm     5.8$ \\
16491.0 & {\rm UVM2}  &  885.6 & $17.493 \pm 0.052$ & $    67.0\pm     3.1$ \\
30028.0 & {\rm UVM2}  &  611.2 & $18.381 \pm 0.101$ & $    29.6\pm     2.6$ \\
47376.0 & {\rm UVM2}  &  610.0 & $19.039 \pm 0.144$ & $    16.1\pm     2.0$ \\
  631.0 & {\rm U}     &   19.4 & $12.474 \pm 0.028$\tablenotemark{x} & $ 19697.0\pm   501.5$\tablenotemark{x} \\
  785.0 & {\rm U}     &   19.4 & $12.968 \pm 0.030$ & $ 12496.8\pm   340.6$ \\
 1430.0 & {\rm U}     &   19.4 & $14.108 \pm 0.042$ & $  4373.2\pm   165.9$ \\
 1591.0 & {\rm U}     &   19.5 & $14.342 \pm 0.046$ & $  3525.3\pm   146.2$ \\
 4926.0 & {\rm U}     &  196.6 & $16.015 \pm 0.029$ & $   755.1\pm    19.9$ \\
 6362.0 & {\rm U}     &  196.6 & $16.329 \pm 0.033$ & $   565.5\pm    16.9$ \\
18304.0 & {\rm U}     &  295.0 & $17.864 \pm 0.056$ & $   137.5\pm     6.9$ \\
18609.0 & {\rm U}     &  231.2 & $17.886 \pm 0.064$ & $   134.8\pm     7.7$ \\
28486.0 & {\rm U}     &  213.4 & $18.498 \pm 0.089$ & $    76.7\pm     6.0$ \\
45833.0 & {\rm U}     &  213.4 & $19.435 \pm 0.148$ & $    32.4\pm     4.1$ \\
  656.0 & {\rm B}     &    9.6 & $13.334 \pm 0.042$\tablenotemark{x} & $ 19160.2\pm   727.0$\tablenotemark{x} \\
  811.0 & {\rm B}     &    9.6 & $13.913 \pm 0.046$ & $ 11240.9\pm   466.3$ \\
 1454.0 & {\rm B}     &   19.4 & $15.012 \pm 0.046$ & $  4085.1\pm   169.5$ \\
 1616.0 & {\rm B}     &   19.5 & $15.291 \pm 0.052$ & $  3159.4\pm   147.7$ \\
 5131.0 & {\rm B}     &  196.6 & $16.962 \pm 0.033$ & $   678.0\pm    20.3$ \\
 6567.0 & {\rm B}     &  196.6 & $17.350 \pm 0.039$ & $   474.2\pm    16.7$ \\
28708.0 & {\rm B}     &  213.4 & $19.417 \pm 0.112$ & $    70.7\pm     6.9$ \\
46055.0 & {\rm B}     &  213.3 & $20.036 \pm 0.171$ & $    40.0\pm     5.8$ \\
75943.0 & {\rm B}     &  100.2 & $20.441 \pm 0.316$ & $    27.5\pm     6.9$ \\
  711.0 & {\rm V}     &   19.5 & $13.373 \pm 0.040$\tablenotemark{x} & $ 16512.0\pm   597.2$\tablenotemark{x} \\
  975.0 & {\rm V}     &  393.5 & $14.337 \pm 0.013$\tablenotemark{x} & $  6795.2\pm    80.9$\tablenotemark{x} \\
 1518.0 & {\rm V}     &   19.4 & $15.025 \pm 0.076$ & $  3605.8\pm   243.8$ \\
 1680.0 & {\rm V}     &   19.5 & $15.127 \pm 0.081$ & $  3282.5\pm   236.0$ \\
 5746.0 & {\rm V}     &  196.6 & $16.955 \pm 0.058$ & $   609.5\pm    31.7$ \\
11616.0 & {\rm V}     &  295.0 & $17.948 \pm 0.078$ & $   244.2\pm    16.9$ \\
11919.0 & {\rm V}     &  295.1 & $17.895 \pm 0.078$ & $   256.4\pm    17.8$ \\
12222.0 & {\rm V}     &  295.0 & $18.041 \pm 0.082$ & $   224.2\pm    16.3$ \\
29806.0 & {\rm V}     &  213.4 & $19.491 \pm 0.256$ & $    59.0\pm    12.4$ \\
\enddata 
\tablenotetext{a}{Exposure mid-time, measured from the \emph{Swift} trigger (UTC 06:12:49).}
\tablenotetext{b}{Observed value; not corrected for Galactic extinction.}
\tablenotetext{x}{Point not used in modeling.}
\end{deluxetable}

\begin{deluxetable}{rccll}
\tabletypesize{\small}
\tablecaption{Gemini Observations of GRB\,080319B\label{tab:photloggemini}}
\tablecolumns{6}
\tablehead{
\colhead{$t_{\rm mid}$\tablenotemark{a}} & \colhead{Filter} &
\colhead{Exp.~time} &
\colhead{Mag.\tablenotemark{b}} & \colhead{Flux\tablenotemark{b}} \\
\colhead{sec} & \colhead{} &
\colhead{sec} & \colhead{} &
\colhead{$\mu$Jy}}
\startdata
       89309 & {\rm g} & 5 $\times$ 180 & $20.670 \pm 0.100$ & $   19.59\pm    1.72$ \\
       90356 & {\rm r} & 5 $\times$ 180 & $20.520 \pm 0.060$ & $   22.49\pm    1.21$ \\
       91307 & {\rm i} & 5 $\times$ 180 & $20.380 \pm 0.050$ & $   25.59\pm    1.15$ \\
       92470 & {\rm z} & 5 $\times$ 180 & $20.310 \pm 0.050$ & $   27.93\pm    1.26$ \\
      174960 & {\rm r} & 5 $\times$ 180 & $21.510 \pm 0.060$ & $   9.037\pm   0.486$ \\
      520390 & {\rm g} & 6 $\times$ 180 & $23.450 \pm 0.190$ & $   1.514\pm   0.243$ \\
      521589 & {\rm r} & 6 $\times$ 180 & $23.340 \pm 0.090$ & $   1.675\pm   0.133$ \\
      522792 & {\rm i} & 6 $\times$ 180 & $22.900 \pm 0.060$ & $   2.512\pm   0.135$ \\
      523990 & {\rm z} & 6 $\times$ 180 & $22.940 \pm 0.090$ & $   2.477\pm   0.197$ \\
      780120 & {\rm r} & 6 $\times$ 180 & $23.670 \pm 0.070$ & $   1.236\pm   0.077$ \\
      781560 & {\rm i} & 6 $\times$ 180 & $23.280 \pm 0.060$ & $   1.770\pm   0.095$ \\
     1209960 & {\rm g} & 6 $\times$ 180 & $24.440 \pm 0.080$ & $  0.6081\pm  0.0432$ \\
     1211220 & {\rm r} & 6 $\times$ 180 & $24.100 \pm 0.080$ & $  0.8318\pm  0.0591$ \\
     1212480 & {\rm i} & 6 $\times$ 180 & $23.810 \pm 0.070$ & $   1.086\pm   0.068$ \\
     2286720 & {\rm g} & 6 $\times$ 180 & $25.860 \pm 0.110$\tablenotemark{x} & $  0.1644\pm  0.0158$\tablenotemark{x} \\
     2287800 & {\rm r} & 6 $\times$ 180 & $25.050 \pm 0.090$\tablenotemark{x} & $  0.3467\pm  0.0276$\tablenotemark{x} \\
     2285280 & {\rm i} & 6 $\times$ 180 & $24.360 \pm 0.080$\tablenotemark{x} & $  0.6546\pm  0.0465$\tablenotemark{x} \\
     2531160 & {\rm z} & 3 $\times$ 180 & $23.940 \pm 0.130$\tablenotemark{x} & $  0.9863\pm  0.1113$\tablenotemark{x} \\
\enddata 
\tablenotetext{a}{Exposure mid-time, measured from the \emph{Swift} trigger (UTC 06:12:49).}
\tablenotetext{b}{Observed value; not corrected for Galactic extinction.}
\tablenotetext{x}{Point not used in modeling.}
\end{deluxetable}

\begin{deluxetable}{rccllll}
\tabletypesize{\small}
\tablecaption{GCN Observations of GRB\,080319B\label{tab:photloggcn}}
\tablecolumns{7}
\tablehead{
\colhead{$t_{\rm mid}$\tablenotemark{a}} & \colhead{Filter} &
\colhead{Exp.~time} &
\colhead{Mag.\tablenotemark{b}} & \colhead{Flux\tablenotemark{b}} & \colhead{} & \colhead{} \\
\colhead{sec} & \colhead{} &
\colhead{sec} & \colhead{} & 
\colhead{$\mu$Jy} & \colhead{} & \colhead{} }
\startdata
  -11.0 & {\rm clear} &   10.0 & $ \geq 11.48$ & $ \leq7.94\times 10^{4}$ & Pi\tablenotemark{c} & GCN 7445\tablenotemark{d}\\
    3.0 & {\rm clear} &   10.0 & $  9.83$ & $3.63\times 10^{5}$ & Pi & GCN 7445\\
   17.0 & {\rm clear} &   10.0 & $  5.76$ & $1.54\times 10^{7}$ & Pi & GCN 7445\\
   32.0 & {\rm clear} &   10.0 & $  6.00$ & $1.24\times 10^{7}$ & Pi & GCN 7445\\
   79.0 & {\rm clear} &   10.0 & $  8.26$ & $1.54\times 10^{6}$ & Pi & GCN 7445\\
   93.0 & {\rm clear} &   10.0 & $  8.77$ & $9.64\times 10^{5}$ & Pi & GCN 7445\\
  108.0 & {\rm clear} &   10.0 & $  9.10$ & $7.11\times 10^{5}$ & Pi & GCN 7445\\
  122.0 & {\rm clear} &   10.0 & $ 10.27$ & $2.42\times 10^{5}$ & Pi & GCN 7445\\
  163.0 & {\rm clear} &   10.0 & $ 10.50$ & $1.96\times 10^{5}$ & Pi & GCN 7445\\
  177.0 & {\rm clear} &   10.0 & $ 11.10$ & $1.13\times 10^{5}$ & Pi & GCN 7445\\
  252.0 & {\rm clear} &   10.0 & $ 11.21$ & $1.02\times 10^{5}$ & Pi & GCN 7445\\
  296.0 & {\rm clear} &   10.0 & $ 11.79$ & $5.97\times 10^{4}$ & Pi & GCN 7445\\
  310.0 & {\rm clear} &   10.0 & $ 11.95$ & $5.15\times 10^{4}$ & Pi & GCN 7445\\
   37.2 & {\rm clear} &   10.0 & $  5.60$ & $1.79\times 10^{7}$ & RAPTOR & GCN 7464\tablenotemark{e} \\
   24.6 & {\rm clear} &    5.0 & $  5.35$ & $2.25\times 10^{7}$ & ROTSE & GCN 7470\tablenotemark{f}\\
  111.7 & {\rm clear} &   20.0 & $  8.49$ & $1.25\times 10^{6}$ & ROTSE & GCN 7470\\
    8.9 & {\rm clear}\tablenotemark{v}     &      -- & $ 8.198 \pm 0.579$ & $1.94\times 10^{6}\pm8.02\times 10^{5}$ & TORTORA & GCN 7558\tablenotemark{g} \\
   10.1 & {\rm clear}\tablenotemark{v}     &      -- & $ 6.967 \pm 0.188$ & $6.03\times 10^{6}\pm9.58\times 10^{5}$ & TORTORA & GCN 7558\\
   11.5 & {\rm clear}\tablenotemark{v}     &      -- & $ 6.650 \pm 0.141$ & $8.07\times 10^{6}\pm9.83\times 10^{5}$ & TORTORA & GCN 7558\\
60300.0 & {\rm R}     &      -- & $ 19.80$           & $3.73\times 10^{1}$                     & Canarias  & GCN 7476\tablenotemark{h} \\
55830.0 & {\rm clear} & 1500.0 &  $19.080 \pm 0.020$ & $7.24\times 10^{1}\pm1.32\times 10^{0}$ & AAVSO & GCN 7484\tablenotemark{i} \\
70500.0 & {\rm clear} & 2040.0 &  $19.000 \pm 0.100$ & $7.80\times 10^{1}\pm6.86\times 10^{0}$ & AAVSO & GCN 7484 \\
74831.0 & {\rm R}     &      -- & $20.000 \pm 0.300$ & $3.10\times 10^{1}\pm7.50\times 10^{0}$ & Brno  & GCN 7504\tablenotemark{j} \\
43934.4 & {\rm R}     &      -- & $ 19.10$           & $7.11\times 10^{1}$                     & Kharkiv & GCN 7519\tablenotemark{k}  \\
46270.8 & {\rm R}     &      -- & $ 19.90$           & $3.40\times 10^{1}$                     & Kharkiv & GCN 7519   \\
71208.0 & {\rm R}     &      -- & $ 20.10$           & $2.83\times 10^{1}$                     & Kharkiv & GCN 7519   \\
1656000 & {\rm R}\tablenotemark{n}& -- & $ 24.35$           & $5.65\times 10^{-1}$                     & HST & GCN 7569\tablenotemark{l}\\
1656000 & {\rm I}\tablenotemark{o}& -- & $ 23.76$           & $7.62\times 10^{-1}$                     & HST & GCN 7569\\
4590000 & {\rm R}\tablenotemark{n}& -- & $26.280 \pm 0.100$ & $9.55\times 10^{-2}\pm8.40\times 10^{-3}$ & HST & GCN 7710\tablenotemark{m} \\
4590000 & {\rm I}\tablenotemark{o}& -- & $25.460 \pm 0.100$ & $1.59\times 10^{-1}\pm1.40\times 10^{-2}$ & HST & GCN 7710\\
\enddata 
\tablecomments{Most of the TORTORA measurements have been omitted to save space.  The full table of GCN photometry used in our plot is available online.}
\tablenotetext{a}{Exposure mid-time, measured from the {\textit{Swift}}\, trigger (UTC 06:12:49).}
\tablenotetext{b}{Observed value; not corrected for Galactic extinction.}
\tablenotetext{c}{``Pi$-$of$-$the$-$Sky'' Collaboration}
\tablenotetext{d}{\cite{Cwiok08_GCN7445}}
\tablenotetext{e}{\cite{Wozniak08_GCN7464}}
\tablenotetext{f}{\cite{Swan08_GCN7470}}
\tablenotetext{g}{\cite{Karpov08_GCN7558}}
\tablenotetext{h}{\cite{Jelinek08_GCN7476}}
\tablenotetext{i}{\cite{Hentunen08_GCN7484}}
\tablenotetext{j}{\cite{Novak08_GCN7504}}
\tablenotetext{k}{\cite{Krugly08_GCN7519}}
\tablenotetext{l}{\cite{Tanvir08_GCN7569}}
\tablenotetext{m}{\cite{Levan08_GCN7710}}
\tablenotetext{n}{Converted from F606W.}
\tablenotetext{o}{Converted from F804W.}
\tablenotetext{v}{Calibrated to $V$-band and plotted as $V$ in our figures.}
\tablenotetext{x}{No GCN observations were used in modeling.}
\end{deluxetable}

\begin{deluxetable}{llllll}
\tabletypesize{\small}
\tablecaption{Bright SDSS standard stars}
\tablecolumns{6}
\tablehead{
\colhead{RA} & \colhead{Dec} & \colhead{B} & \colhead{V} & \colhead{R} & \colhead{I}  \\
\colhead{deg} & \colhead{deg} & \colhead{mag} & \colhead{mag} & \colhead{mag} & \colhead{mag}
}
\startdata
217.915229 & 36.31513  & 15.940 $\pm$ 0.010 & 15.293 $\pm$ 0.006 &14.922 $\pm$ 0.008 & 14.551 $\pm$ 0.007 \\
217.908143 & 36.29125  & 17.716 $\pm$ 0.012 & 16.350 $\pm$ 0.006 &15.523 $\pm$ 0.008 & 14.821 $\pm$ 0.007 \\
217.943617 & 36.31667  & 17.993 $\pm$ 0.012 & 16.834 $\pm$ 0.006 &16.162 $\pm$ 0.008 & 15.590 $\pm$ 0.007 \\
\enddata 
\label{tab:lickstars}
\end{deluxetable}

\begin{deluxetable}{llllll}
\tabletypesize{\small}
\tablecaption{Faint SDSS standard stars}
\tablecolumns{6}
\tablehead{
\colhead{RA} & \colhead{Dec} & \colhead{g} & \colhead{r} & \colhead{i} & \colhead{z}  \\
\colhead{deg} & \colhead{deg} & \colhead{mag} & \colhead{mag} & \colhead{mag} & \colhead{mag}
}
\startdata
217.908143 & 36.29125 & 17.091 $\pm$ 0.005 & 15.816 $\pm$ 0.004 & 15.310 $\pm$ 0.004 & 15.067 $\pm$ 0.005 \\
217.943617 & 36.31667 & 17.425 $\pm$ 0.005 & 16.409 $\pm$ 0.004 & 16.058 $\pm$ 0.004 & 15.872 $\pm$ 0.007 \\
217.871828 & 36.29380 & 17.364 $\pm$ 0.005 & 16.903 $\pm$ 0.005 & 16.769 $\pm$ 0.005 & 16.689 $\pm$ 0.010 \\
217.894280 & 36.34851 & 18.298 $\pm$ 0.007 & 16.955 $\pm$ 0.005 & 16.370 $\pm$ 0.005 & 16.036 $\pm$ 0.007 \\
217.902832 & 36.33332 & 18.084 $\pm$ 0.006 & 17.741 $\pm$ 0.006 & 17.626 $\pm$ 0.007 & 17.565 $\pm$ 0.017 \\
217.981086 & 36.29605 & 18.223 $\pm$ 0.007 & 17.866 $\pm$ 0.006 & 17.746 $\pm$ 0.008 & 17.723 $\pm$ 0.041 \\
217.931339 & 36.34208 & 19.812 $\pm$ 0.014 & 18.342 $\pm$ 0.008 & 16.818 $\pm$ 0.005 & 16.016 $\pm$ 0.007 \\
217.931216 & 36.27044 & 19.172 $\pm$ 0.010 & 18.382 $\pm$ 0.008 & 18.110 $\pm$ 0.009 & 17.988 $\pm$ 0.021 \\
217.918619 & 36.31995 & 20.742 $\pm$ 0.026 & 19.264 $\pm$ 0.013 & 18.483 $\pm$ 0.011 & 18.055 $\pm$ 0.024 \\
217.913905 & 36.28612 & 20.736 $\pm$ 0.027 & 19.266 $\pm$ 0.013 & 18.253 $\pm$ 0.010 & 17.687 $\pm$ 0.019 \\
217.928915 & 36.33948 & 20.850 $\pm$ 0.029 & 19.462 $\pm$ 0.017 & 18.214 $\pm$ 0.010 & 17.578 $\pm$ 0.018 \\
217.928801 & 36.30674 & 20.844 $\pm$ 0.028 & 19.471 $\pm$ 0.014 & 18.689 $\pm$ 0.012 & 18.315 $\pm$ 0.029 \\
217.955190 & 36.33084 & 19.708 $\pm$ 0.014 & 19.506 $\pm$ 0.015 & 19.505 $\pm$ 0.023 & 19.315 $\pm$ 0.065 \\
217.926000 & 36.33788 & 20.047 $\pm$ 0.016 & 19.640 $\pm$ 0.016 & 19.474 $\pm$ 0.022 & 19.437 $\pm$ 0.071 \\
217.932586 & 36.27256 & 21.248 $\pm$ 0.033 & 19.710 $\pm$ 0.016 & 18.342 $\pm$ 0.010 & 17.551 $\pm$ 0.016 \\
217.946853 & 36.30972 & 21.362 $\pm$ 0.040 & 19.909 $\pm$ 0.020 & 19.003 $\pm$ 0.016 & 18.439 $\pm$ 0.033 \\
217.971356 & 36.30059 & 21.318 $\pm$ 0.039 & 19.937 $\pm$ 0.020 & 19.043 $\pm$ 0.016 & 18.571 $\pm$ 0.035 \\
\enddata
\label{tab:sdssstars}
\end{deluxetable}
\end{document}